\documentclass[preprint,12pt]{elsarticle}

\usepackage{amssymb}
\journal{XXX}

\begin{document}

\begin{frontmatter}



\title{An efficient method to construct self-dual cyclic codes of length $p^s$
over $\mathbb{F}_{p^m}+u\mathbb{F}_{p^m}$}


\author{Yuan Cao$^{a, \ b, \ c}$, Yonglin Cao$^{a, \ \ast}$, Hai Q. Dinh$^{d, \ e}$, Somphong Jitman$^{f}$}

\address{$^{a}$School of Mathematics and Statistics,
Shandong University of Technology, Zibo, Shandong 255091, China

\vskip 1mm $^{b}$Hubei Key Laboratory of Applied Mathematics, Faculty of Mathematics and Statistics, Hubei University, Wuhan 430062, China

\vskip 1mm $^{c}$Hunan Provincial Key Laboratory of Mathematical Modeling and Analysis in Engineering, Changsha University of Science and Technology, Changsha, Hunan 410114, China

\vskip 1mm $^{d}$Division of Computational Mathematics and Engineering, Institute for Computational
       Science, Ton Duc Thang University, Ho Chi Minh City, Vietnam

\vskip 1mm $^{e}$Faculty of Mathematics and Statistics, Ton Duc Thang University, Ho Chi Minh City,
      Vietnam

\vskip 1mm $^{f}$Department of Mathematics, Faculty of Science, Silpakorn University, Nakhon
Pathom 73000, Thailand}
\cortext[cor1]{corresponding author.  \\
E-mail addresses: yuancao@sdut.edu.cn (Yuan Cao), \ ylcao@sdut.edu.cn (Yonglin Cao),
\ dinhquanghai@tdtu.edu.vn (H. Q. Dinh), \ sjitman@gmail.com (S. Jitman).}

\begin{abstract}
Let $p$ be an odd prime number, $\mathbb{F}_{p^m}$ be a finite field of cardinality $p^m$ and $s$ a positive integer.
Using some combinatorial identities, we obtain certain properties for Kronecker product of matrices over $\mathbb{F}_p$ with a specific type.
On that basis, we give an explicit representation and enumeration for all distinct self-dual cyclic codes of length $p^s$ over the finite chain ring $\mathbb{F}_{p^m}+u\mathbb{F}_{p^m}$ $(u^2=0)$. Moreover,
We provide
an efficient method to construct every self-dual cyclic code of length $p^s$ over  $\mathbb{F}_{p^m}+u\mathbb{F}_{p^m}$
precisely.
\end{abstract}


\end{frontmatter}


\section{Introduction}

\noindent
  The class of self-dual codes is an interesting topic in coding theory due to
their connections to other fields of mathematics such as Lattices, Cryptography, Invariant Theory, Block designs, etc.
An effective way for the construction of self-dual codes is the use of some specific algebraic structures.

\par
   Let $\mathbb{F}_{p^m}$ be a finite field of $p^m$ elements, where $p$ is a prime number, and denote
$R=\frac{\mathbb{F}_{p^m}[u]}{\langle u^2\rangle}=\mathbb{F}_{p^m}
+u\mathbb{F}_{p^m} \ (u^2=0).$
Then $R$ is a finite chain ring and every invertible element in $R$ is of the form: $a+bu$, $a,b\in \mathbb{F}_{p^m}$ and $a\neq 0$.
Let $N$ be a fixed positive integer and
$R^N=\{(a_0,a_1,\ldots,a_{N-1})\mid a_0,a_1,\ldots,a_{N-1}\in R\}$
Then $R^N$ is an $R$-free module with the usual componentwise addition and scalar multiplication by elements of $R$.
Let $\mathcal{C}$ be an
$R$-submodule of $R^N$ and $\lambda$ be an invertible element in $R$. Then $\mathcal{C}$ is called a \textit{linear code} over
$R$ of length $N$. Moreover, $\mathcal{C}$ is called a \textit{$\lambda$-constacyclic code} if
$$(\lambda a_{N-1},a_0,a_1,\ldots,a_{N-2})\in \mathcal{C}, \ \forall (a_0,\ldots,a_{N-2},a_{N-1})\in \mathcal{C}.$$
In particular, a $\lambda$-constacyclic code $\mathcal{C}$ is called a \textit{negacyclic code} when $\lambda=-1$, and
$\mathcal{C}$ is called a \textit{cyclic code} when $\lambda=1$.

\par
   Let $\frac{R[x]}{\langle x^N-\lambda\rangle}=\{\sum_{i=0}^{N-1}a_ix^i\mid a_0,a_1,\ldots,a_{N-1}\in R\}$
in which the arithmetic is done modulo $x^N-\lambda$.
In this paper, $\lambda$-constacyclic codes over
$R$ of length $N$ are identified with ideals of the ring $\frac{R[x]}{\langle x^N-\lambda\rangle}$, under the
identification map $\theta: R^N \rightarrow \frac{R[x]}{\langle x^N-\lambda\rangle}$ defined by
$\theta: (a_0,a_1,\ldots,a_{N-1})\mapsto
a_0+a_1x+\ldots+a_{N-1}x^{N-1}$ for all $a_i\in R$ and $i=0,1,\ldots,N-1$.

\par
  The \textit{Euclidean inner
product} on $R^N$ is defined by
$[\alpha,\beta]=\sum_{i=0}^{N-1}a_ib_i\in R$
for
all $\alpha=(a_0,a_1,\ldots,a_{N-1}), \beta=(b_0,b_1,\ldots,b_{N-1})\in R^N$. Then
the (Euclidean) \textit{dual code} of a linear code $\mathcal{C}$ over $R$ of length $N$ is defined by
$$\mathcal{C}^{\bot}=\{\beta\in R^N\mid [\alpha,\beta]=0, \ \forall \alpha\in \mathcal{C}\},$$
which is also a linear code over $R$ of length $N$. In particular, $\mathcal{C}$ is said to be
 (Euclidean) \textit{self-dual} if $\mathcal{C}^{\bot}=\mathcal{C}$.

\par
  There were a lot of literature on linear codes, cyclic codes and
constacyclic codes of length $N$ over rings $\mathbb{F}_{p^m}+u\mathbb{F}_{p^m}$ ($u^2=0$) for various prime $p$, positive integer $m$ and some positive integer $N$ (see \cite{s1}, \cite{s2}, \cite{s3}, \cite{s4} and \cite{s6}--\cite{s18}, for examples).

\par
  Specifically,
all constacyclic codes of length $2^s$ over the Galois extension
rings of $\mathbb{F}_2 + u\mathbb{F}_2$ were classified and their detailed structures was also established in \cite{s9}. Dinh \cite{s10}
classified all constacyclic codes of length $p^s$ over $\mathbb{F}_{p^m}+u\mathbb{F}_{p^m}$. Then
negacyclic codes of length $2p^s$, constacyclic codes of length $2p^s$ and
constacyclic codes of length $4p^s$ ($p^m\equiv 1$ (mod $4$)) over $\mathbb{F}_{p^m}+u\mathbb{F}_{p^m}$
were investigated by Dinh et al. \cite{s11}, Chen et al. \cite{s8} and Dinh et al. \cite{s12}, respectively.
We note that \textsl{the representation and enumeration for self-dual cyclic codes and self-dual negacyclic codes were not studied in these papers}.

\par
   Dinh et al. \cite{s13} determined the algebraic structures of all cyclic and negacyclic codes
of length $4p^s$ over $\mathbb{F}_{p^m}+u\mathbb{F}_{p^m}$, established the duals of all such codes and gave some special subclass of self-dual negacyclic codes of length $4p^s$ over $\mathbb{F}_{p^m}+u\mathbb{F}_{p^m}$ by Theorems 4.2, 4.4
and 4.9 of \cite{s13}. But \textsl{the representation and enumeration for all self-dual negacyclic codes and all self-dual cyclic codes
were not given}.

\par
  Choosuwan et al. \cite{s5} done the following:

\par
  $\diamondsuit$ In pages $9$ and $10$, they proved that every
(Euclidean) self-dual cyclic code over $\mathbb{F}_{p^m}+u\mathbb{F}_{p^m}$
of length $p^s$ is given by
$$\mathcal{C}=\langle (x-1)^{i_0}+u\sum_{j=0}^{i_1-1}h_j(x-1)^j,(x-1)^{i_1}\rangle,$$
where $p^s=i_0+i_1$ and $\textbf{h}=(h_0,h_1,\ldots,h_{i_1-1})^{{\rm tr}}\in \mathbb{F}_{p^m}^{i_1}$
satisfying
$$M(p^s,i_1)\textbf{h}=\textbf{0}$$
in which $M(p^s,i_1)$ is an $i_1\times i_1$ matrix over $\mathbb{F}_{p^m}$ defined by
{\tiny
$$M(p^s,i_1)=\left(\begin{array}{ccccc}
(-1)^{i_0}+1 & 0 & 0 & \ldots & 0  \cr
(-1)^{i_0}\left(\begin{array}{c}i_0 \cr 1\end{array}\right) & (-1)^{i_0+1}+1 & 0 & \ldots & 0 \cr
(-1)^{i_0}\left(\begin{array}{c}i_0 \cr 2\end{array}\right) & (-1)^{i_0+1}\left(\begin{array}{c}i_0 \cr 1\end{array}\right)
& (-1)^{i_0+2}+1 & \ldots & 0 \cr
\ldots & \ldots & \ldots & \ldots & \ldots \cr
(-1)^{i_0}\left(\begin{array}{c}i_0 \cr i_1-1\end{array}\right) & (-1)^{i_0+1}\left(\begin{array}{c}i_0-1 \cr i_1-2\end{array}\right)
& (-1)^{i_0+2}\left(\begin{array}{c}i_0-2 \cr i_1-3\end{array}\right) & \ldots & (-1)^{i_0+i_1-1}+1
\end{array}\right).$$ }

\par
  $\diamondsuit$ Using Theorem 3.3 of \cite{s19} for the nullity of
$M(p^s,i_1)$, they obtained a formula to
count the number of self-dual cyclic codes over $\mathbb{F}_{p^m}+u\mathbb{F}_{p^m}$
of length $p^s$ (cf. Corollary 22 of \cite{s5}), where $p$ is an arbitrary prime number.

\par
  Also Dinh et al. determined
the number of self-dual cyclic codes of length $p^s$ over $\mathbb{F}_{p^m}+u\mathbb{F}_{p^m}$
($u^2=0$) by Section 4 of \cite{DM2018}.

\par
  But they didn't give a method how to solve the equation
$M(p^s,i_1)\textbf{h}=\textbf{0}$ and didn't obtain an representation for
solutions of this equation in \cite{s5} and the equation (2.1) in \cite{DM2018}. So they didn't provide an explicit
representation for all distinct self-dual cyclic codes over $\mathbb{F}_{p^m}+u\mathbb{F}_{p^m}$
of length $p^s$.

\par
   In \cite{s6}, we provided a new way different
from the methods used in \cite{s8}--\cite{s14} to determine the algebraic structures,
generators and enumeration of $\lambda$-constacyclic codes over $\mathbb{F}_{p^m}+u\mathbb{F}_{p^m}$ of length
$np^s$, where $n$ is an arbitrary positive integer satisfying ${\rm gcd}(p,n)=1$ and $\lambda\in \mathbb{F}_{p^m}^\times$.
Then we gave an explicit representation for the dual code of  every cyclic
code and every negacyclic code. Moreover, we provided a discriminant condition for the self-duality
of each cyclic code and negacyclic code over $\mathbb{F}_{p^m}+u\mathbb{F}_{p^m}$ of length
$np^s$. On the basis of \cite{s6}, we can consider to give an explicit representation for self-dual cyclic
codes and self-dual negacyclic codes over $\mathbb{F}_{p^m}+u\mathbb{F}_{p^m}$.

\par
  Recently,
by a new way different from that of \cite{s5},
we \cite{s7} gave an efficient method for the construction of all distinct self-dual cyclic codes with length $2^s$ over  $\mathbb{F}_{2^m}+u\mathbb{F}_{2^m}$. In particular, we provide an exact formula to count the number of all these self-dual cyclic codes
and corrected a mistake in Corollary 22(ii) of \cite{s5}. However, the methods and results of \cite{s7} depend heavily on that
the characteristic of the field $\mathbb{F}_{2^m}$ is $2$.
They can't be used directly to the case for self-dual cyclic codes with length $p^s$ over $\mathbb{F}_{p^m}+u\mathbb{F}_{p^m}$
where $p$ is odd. Hence we need to develop a new approach to the latter situation.

\vskip 2mm\par
  The present paper is organized as follows. In Section 2, we review the
known results for self-dual cyclic codes of length $p^s$ over $\mathbb{F}_{p^m}+u\mathbb{F}_{p^m}$ and prove that these self-dual cyclic codes are determined by a special kind of subsets $\Omega_l$ in the
residue class ring $\frac{\mathbb{F}_{p^m}[x]}{\langle (x-1)^l\rangle}$ for certain integers $l$, $1\leq l\leq p^s-1$. In Section 3,
we give an explicit representation of the set $\Omega_l$ by studying properties for
Kronecker product of matrices over $\mathbb{F}_{p}$ with a specific type. In Section 4, we provide an efficient method to construct
and represent all distinct self-dual cyclic codes of length $p^s$ over $\mathbb{F}_{p^m}+u\mathbb{F}_{p^m}$
precisely. As an application, we list all distinct self-dual cyclic codes over $\mathbb{F}_{3^m}+u\mathbb{F}_{3^m}$
of length $3^s$ for $s=1,2,3$ in Section 5.
Section 6 concludes the paper.

%%%%%%%%%%%%%%%%%%%%%%%%%%%%%%%%%%%%%%%%%%%%%%%%%%%%%%%%%%%%%%%%%%%%%%%

%%%%%%%%%%%%%%%%%%%%%%%%%%%%%%%%%%%%%%%%%%%%%%%%%%%%%%%%%%%%%%%%%%%%%%
%%%%%%%%%%%%%%%%%%%%%%%%%%%%%%%%%%%%%%%%%%%%%%%%%%%%%%%%%%%%%%%%%%%%%%%

%%%%%%%%%%%%%%%%%%%%%%%%%%%%%%%%%%%%%%%%%%%%%%%%%%%%%%%%%%%%%%%%%%%%%%

\section{Preliminaries}

\noindent \par
In this section, we list some known results for cyclic codes  of length $p^s$ over the ring $\mathbb{F}_{p^m}+u\mathbb{F}_{p^m}$ ($u^2=0$)
needed in the following sections.

\par
  By Corollary 7.1 in \cite{s6}, every cyclic code $\mathcal{C}$ over $\mathbb{F}_{p^m}+u\mathbb{F}_{p^m}$ of length
$p^s$ and its dual code $\mathcal{C}^{\bot}$ are
given by the following five cases.
\begin{description}
\item{Case I.}
  $(p^m)^{p^s-\lceil \frac{p^s}{2}\rceil}=p^{\frac{p^s-1}{2}m}$ codes:

\par
  $\mathcal{C}=\langle (x-1)b(x)+u\rangle$ with $|\mathcal{C}|=p^{p^sm}$ and
  $\mathcal{C}^{\bot}=\langle (x-1)\cdot x^{-1}b(x^{-1})+u\rangle$,
where $b(x)\in (x-1)^{\frac{p^s-1}{2}}\cdot \frac{\mathbb{F}_{p^m}[x]}{\langle (x-1)^{p^{s}-1}\rangle}$.

\item{Case II.}
  $\sum_{k=1}^{p^s-1}p^{(p^s-k-\lceil\frac{1}{2}(p^s-k)\rceil)m}$ codes:

\par
  $\mathcal{C}=\langle (x-1)^{k+1}b(x)+u(x-1)^{k}\rangle$ with $|\mathcal{C}|=p^{(p^s-k)m}$
  and
  $$\mathcal{C}^{\bot}=\langle (x-1)\cdot x^{-1}b(x^{-1})+u, (x-1)^{p^s-k}\rangle,$$
where $b(x)\in (x-1)^{\lceil\frac{p^s-k}{2}\rceil-1}\cdot \frac{\mathbb{F}_{p^m}[x]}{\langle (x-1)^{p^{s}-k-1}\rangle}$
and $1\leq k\leq p^s-1$.

\item{Case III.}
   $p^s+1$ codes:

\par
  $\mathcal{C}=\langle (x-1)^k\rangle$ with $|\mathcal{C}|=p^{2(p^s-k)m}$ and $\mathcal{C}^{\bot}=\langle (x-1)^{p^s-k}\rangle$,
where $0\leq k\leq p^s$.

\item{Case IV.}
  $\sum_{t=1}^{p^s-1}p^{(t-\lceil\frac{t}{2}\rceil)m}$ codes:

\par
  $\mathcal{C}=\langle (x-1)b(x)+u,(x-1)^t\rangle$ with $|\mathcal{C}|=p^{(2\cdot p^s-t)m}$ \\
   and $\mathcal{C}^{\bot}=\langle (x-1)^{p^s-t+1}\cdot x^{-1}b(x^{-1})+u(x-1)^{p^s-t}\rangle$, \\
where $b(x)\in (x-1)^{\lceil\frac{t}{2}\rceil-1}\cdot \frac{\mathbb{F}_{p^m}[x]}{\langle (x-1)^{t-1}\rangle}$
and $1\leq t\leq p^s-1$.

\item{Case V.}
  $\sum_{k=1}^{p^s-2}\sum_{t=1}^{p^s-k-1}p^{(t-\lceil\frac{t}{2}\rceil)m}$ codes:

\par
  $\mathcal{C}=\langle (x-1)^{k+1}b(x)+u (x-1)^k,(x-1)^{k+t}\rangle$ with $|\mathcal{C}|=p^{(2\cdot p^s-2k-t)m}$ \\
and $\mathcal{C}^{\bot}=\langle (x-1)^{p^s-k-t+1}\cdot x^{-1}b(x^{-1})+u(x-1)^{p^s-k-t},(x-1)^{p^s-k}\rangle$, \\
where $b(x)\in (x-1)^{\lceil\frac{t}{2}\rceil-1}\cdot \frac{\mathbb{F}_{p^m}[x]}{\langle (x-1)^{t-1}\rangle}$,
$1\leq t\leq p^s-k-1$ and $1\leq k\leq p^s-2$.
\end{description}

\par
  As $|(\mathbb{F}_{p^m}+u\mathbb{F}_{p^m})^{p^s}|=(p^{p^s m})^2$, every self-dual cyclic
code $\mathcal{C}$ over $\mathbb{F}_{p^m}+u\mathbb{F}_{p^m}$ of length $p^s$ must contain
$|\mathcal{C}|=p^{p^s m}$ codewords. From this, we deduce that there is no self-dual codes in
Cases II, III and IV.

\par
  Let  $\mathcal{C}=\langle (x-1)b(x)+u\rangle$ be a code in Case I. Then $\mathcal{C}=\mathcal{C}^{\bot}$ if and only if
$b(x)\in (x-1)^{\frac{p^s-1}{2}}\cdot \frac{\mathbb{F}_{p^m}[x]}{\langle (x-1)^{p^{s}-1}\rangle}$ satisfying $b(x)=x^{-1}b(x^{-1})$, i.e.,
$$x^{-1}b(x^{-1})-b(x)\equiv 0 \ ({\rm mod} \ (x-1)^{p^{s}-1}).$$

\par
  Let $\mathcal{C}=\langle (x-1)^{k+1}b(x)+u (x-1)^k,(x-1)^{k+t}\rangle$ be a code in Case V.
Then $\mathcal{C}=\mathcal{C}^{\bot}$ if and only if $2\cdot p^s-2k-t=p^s$ and $b(x)\in (x-1)^{\lceil\frac{t}{2}\rceil-1}\cdot \frac{\mathbb{F}_{2^m}[x]}{\langle (x-1)^{t-1}\rangle}$ satisfying $b(x)=x^{-1}b(x^{-1})$. The latter is equivalent to
$$x^{-1}b(x^{-1})-b(x)\equiv 0 \ ({\rm mod} \ (x-1)^{t-1}).$$
The former is equivalent to that
$$t=p^s-2k, \ \left\lceil\frac{t}{2}\right\rceil=\frac{p^s-2k}{2}=\frac{p^s-1}{2}-k+1 \ {\rm and} \ 1\leq k\leq \frac{p^s-1}{2}.$$

\par
  In the light of the above discussion, we have the following conclusion.

\vskip 3mm\noindent
  {\bf Proposition 1} \textit{For any integer $l$, $1\leq l\leq p^s-1$, we denote}
\begin{equation}
\label{eq1}
\ \Omega_l=\left\{b(x)\in \frac{\mathbb{F}_{p^m}[x]}{\langle (x-1)^l\rangle}
\mid x^{-1}b(x^{-1})-b(x)\equiv 0 \ ({\rm mod} \ (x-1)^l)\right\}.
\end{equation}
\textit{Then all distinct self-dual cyclic codes over $\mathbb{F}_{p^m}+u\mathbb{F}_{p^m}$ of length
$p^s$ are given by the following two cases}:
\begin{description}
\item{(i)}
   \textit{$\langle (x-1)b(x)+u\rangle$, where $b(x)=\sum_{i=\frac{p^s-1}{2}}^{p^s-2}b_i(x-1)^i\in \Omega_{p^s-1}$}.

\item{(ii)}
  \textit{$\langle (x-1)^{k+1}b(x)+u(x-1)^k, (x-1)^{p^s-k}\rangle$, \\
  where
$1\leq k\leq \frac{p^s-1}{2}$ and  $b(x)=\sum_{i=\frac{p^s-1}{2}-k}^{p^s-2k-2}b_i(x-1)^i\in \Omega_{p^s-1-2k}$}.
\end{description}

\par
   In order to present all self-dual cyclic codes over $\mathbb{F}_{p^m}+u\mathbb{F}_{p^m}$ of length
$p^s$ explicitly, by Proposition 1 we need to determine the following subsets of
$\frac{\mathbb{F}_{p^m}[x]}{\langle (x-1)^l\rangle}$:
$$\Omega_{l}, \ {\rm where} \ l=p^s-1-2k \ {\rm and} \ 0\leq k\leq \frac{p^s-1}{2}.$$

\par
  Let $A=(a_{ij})$ and $B$ be matrices over $\mathbb{F}_{p^m}$ of sizes $k\times t$ and $l\times v$ respectively.
The \textit{Kronecker product} of $A$ and $B$ is
defined by $A\otimes B=(a_{ij}B)$ which is a a matrix over $\mathbb{F}_{p^m}$ of size $kl\times tv$.
For any positive integer $\lambda\leq s$, we define a $p^\lambda\times p^\lambda$ lower triangular matrix
$G_{p^\lambda}$ over $\mathbb{F}_p$ as follows
$$G_{p^\lambda}=\left(\begin{array}{cccc} g_{1,1}^{(p^\lambda)} & 0 & \ldots & 0 \cr
g_{2,1}^{(p^\lambda)} & g_{2,2}^{(p^\lambda)} & \ldots & 0\cr
\ldots &\ldots &\ldots &\ldots \cr
g_{p^\lambda,1}^{(p^\lambda)} & g_{p^\lambda,2}^{(p^\lambda)} & \ldots & g_{p^\lambda,p^\lambda}^{(p^\lambda)}\end{array}\right)
\ ({\rm mod} \ p),$$
where
\begin{equation}
\label{eq2}
g_{i,j}^{(p^\lambda)}=(-1)^{j-1}\left(\begin{array}{c}p^\lambda-j\cr i-j\end{array}\right), \
{\rm if} \ 1\leq j\leq i\leq p^\lambda.
\end{equation}
In fact, we have $G_{p^\lambda}=(-1)^{p^s-p^\lambda}M(p^s,p^\lambda)$ (cf. \cite{s5}). Precisely, we have
{\small
$$G_p=\left(\begin{array}{ccccc}\left(\begin{array}{c}p-1\cr 0\end{array}\right) & 0 & 0 & \ldots & 0 \cr
\left(\begin{array}{c}p-1\cr 1\end{array}\right) & (-1)\left(\begin{array}{c}p-2\cr 0\end{array}\right) & 0 & \ldots & 0 \cr
\left(\begin{array}{c}p-1\cr 2\end{array}\right) & (-1)\left(\begin{array}{c}p-2\cr 1\end{array}\right) &
(-1)^2\left(\begin{array}{c}p-3\cr 0\end{array}\right)  & \ldots & 0 \cr
\ldots & \ldots & \ldots & \ldots & \ldots \cr
\left(\begin{array}{c}p-1\cr p-1\end{array}\right) & (-1)\left(\begin{array}{c}p-2\cr p-2\end{array}\right) &
(-1)^2\left(\begin{array}{c}p-3\cr p-3\end{array}\right)  & \ldots & (-1)^{p-1}\left(\begin{array}{c}0\cr 0\end{array}\right)
\end{array}\right)$$}
(mod $p$), where $\left(\begin{array}{c}0\cr 0\end{array}\right)=1$. Moreover, we have the following property for the matrix $G_{p^\lambda}$.

\vskip 3mm\noindent
  {\bf Proposition 2} \textit{Let $\lambda$ be any positive integer and set $G_{p^0}=1$. Then}
\begin{eqnarray*}
G_{p^\lambda}&=&G_p\otimes G_{p^{\lambda-1}}\\
&=&\left(\begin{array}{ccccc}g_{1,1}^{(p)}G_{p^{\lambda-1}} & 0 & 0 & \ldots & 0 \cr
g_{2,1}^{(p)}G_{p^{\lambda-1}}
& g_{2,2}^{(p)}G_{p^{\lambda-1}} & 0 & \ldots & 0 \cr
g_{3,1}^{(p)}G_{p^{\lambda-1}}
& g_{3,2}^{(p)}G_{p^{\lambda-1}} &
g_{3,3}^{(p)}G_{p^{\lambda-1}}  & \ldots & 0 \cr
\ldots & \ldots & \ldots & \ldots & \ldots \cr
g_{p,1}^{(p)}G_{p^{\lambda-1}}
& g_{p,2}^{(p)}G_{p^{\lambda-1}} &
g_{p,3}^{(p)}G_{p^{\lambda-1}}
 & \ldots & g_{p,p}^{(p)}G_{p^{\lambda-1}}
\end{array}\right),
\end{eqnarray*}
\textit{where}
$g_{i,i}^{(p)}=(-1)^{i-1}\left(\begin{array}{c}p-i\cr i-i\end{array}\right)=(-1)^{i-1}, \
i=1,2,\ldots,p.$

\vskip 3mm\noindent
 {\bf Proof.} Let $1\leq j\leq i\leq p$ and $1\leq l\leq k\leq p^{\lambda-1}$, where $\lambda\geq 2$.
As $p^{\lambda-1}$ is odd, we have $(-1)^{(j-1)p^{\lambda-1}+(l-1)}=(-1)^{(j-1)+(l-1)}$. Then by Lucas's Theorem for
a combinatorial identity in number theory
(see \cite{Benjamin}, for examples),
we have
$$\left(\begin{array}{c}(p-j)p^{\lambda-1}+(p^{\lambda-1}-l) \cr (i-j)p^{\lambda-1}+(k-l)\end{array}\right)
\equiv\left(\begin{array}{c} p-j\cr i-j\end{array}\right)
\left(\begin{array}{c} p^{\lambda-1}-l \cr k-l\end{array}\right) \
({\rm mod} \ p).$$
From these, by Equation (\ref{eq2}),
$$p^\lambda-((j-1)p^{\lambda-1}+l)=(p-j)p^{\lambda-1}+(p^{\lambda-1}-l)$$
and $((i-1)p^{\lambda-1}+k)-((j-1)p^{\lambda-1}+l)=(i-j)p^{\lambda-1}+(k-l)$, we deduce
\begin{eqnarray*}
 &&g_{(i-1)p^{\lambda-1}+k,(j-1)p^{\lambda-1}+l}^{(p^\lambda)} \\
 &=& (-1)^{(j-1)p^{\lambda-1}+l-1}\left(\begin{array}{c} p^\lambda-\left((j-1)p^{\lambda-1}+l\right) \cr \left((i-1)p^{\lambda-1}+k\right)-\left((j-1)p^{\lambda-1}+l\right)\end{array}\right)\\
 &=& (-1)^{j-1}\left(\begin{array}{c} p-j\cr i-j\end{array}\right)\cdot (-1)^{l-1}\left(\begin{array}{c} p^{\lambda-1}-l \cr k-l\end{array}\right)\\
 &=&g_{i,j}^{(p)}g_{k,l}^{(p^{\lambda-1})}.
\end{eqnarray*}
This implies that $G_{p^\lambda}=G_p\otimes G_{p^{\lambda-1}}$.
\hfill $\Box$

\vskip 3mm\par
  For examples, we have $G_3=\left(\begin{array}{ccc} 1 & 0 & 0\cr -1 & -1 & 0 \cr 1 & -1 & 1\end{array}\right)$
where $-1=2$, and
{\small
$$G_9=\left(\begin{array}{ccc} G_3 & 0 & 0\cr -G_3 & -G_3 & 0 \cr G_3 & -G_3 & G_3\end{array}\right)
=\left(\begin{array}{ccc|ccc|ccc} 1 & 0 & 0 & 0 & 0 & 0 & 0 & 0 & 0
\cr -1 & -1 & 0 & 0 & 0 & 0 & 0 & 0 & 0 \cr 1 & -1 & 1 & 0 & 0 & 0 & 0 & 0 & 0 \cr\hline
-1 & 0 & 0 & -1 & 0 & 0 & 0 & 0 & 0
\cr 1 & 1 & 0 & 1 & 1 & 0 & 0 & 0 & 0 \cr -1 & 1 & -1 & -1 & 1 & -1 & 0 & 0 & 0
\cr\hline
1 & 0 & 0 & -1 & 0 & 0 & 1 & 0 & 0
\cr -1 & -1 & 0 & 1 & 1 & 0 & -1 & -1 & 0 \cr 1 & -1 & 1 & -1 & 1 & -1 & 1 & -1 & 1\end{array}\right).$$ }

%%%%%%%%%%%%%%%%%%%%%%%%%%%%%%%%%%%%%%%%%%%%%%%%%%%%%%%%%%%%%%%%%%%%%%%%%%%%%%

%%%%%%%%%%%%%%%%%%%%%%%%%%%%%%%%%%%%%%%%%%%%%%%%%%%%%%%%%%%%%%%%%%%%%%%%%%%%%%%%%%%%%%%%%%

%%%%%%%%%%%%%%%%%%%%%%%%%%%%%%%%%%%%%%%%%%%%%%%%%%%%%%%%%%%%%%%%%%%%%%%%%%%%%%%%%%%%%%%%%%%%%%
\section{Calculation and representation of the set $\Omega_l$} \label{}

\noindent
  In this section, we consider how to calculate and represent the subset $\Omega_l$ of
$\frac{\mathbb{F}_{p^m}[x]}{\langle (x-1)^l\rangle}$ defined by Equation (\ref{eq1}), where $1\leq l\leq p^s$.

\par
  For any matrix $A$ over $\mathbb{F}_{p^m}$ and positive integer $l$, let $A^{{\rm tr}}$ be the transpose of $A$
and $I_l$ be the identity matrix of order $l$. In the rest of this paper, we adopt
the following notation
\begin{equation}
\label{eq3}
\mathcal{S}_l=\{B_l=\left(\begin{array}{c}b_0\cr b_1\cr \ldots\cr b_{l-1} \end{array}\right)
\mid \sum_{i=0}^{l-1}b_i(x-1)^i\in \Omega_l, \ b_i\in \mathbb{F}_{p^m}, \ i=0,1,\ldots,l-1\}.
\end{equation}
Then we have
$$\Omega_l=\{(1,(x-1),\ldots,(x-1)^{l-1})B_l\mid B_l\in \mathcal{S}_l\} \ {\rm and} \ |\Omega_l|=|\mathcal{S}_l|.$$
In order to present the
subset $\Omega_l$ of polynomials in $\frac{\mathbb{F}_{p^m}[x]}{\langle (x-1)^l\rangle}$, it is equivalent to determine the set $\mathcal{S}_l$.

\vskip 3mm
\noindent
  {\bf Theorem 1}
\textit{Let $1\leq l\leq p^s-1$ and assume $\lambda$ be the least positive integer
such that $1\leq l\leq p^\lambda.$
 Let $G_l$ be the
submatrix in the upper left corner of
$G_{p^\lambda}$ defined by
\begin{equation}
\label{eq4}
\left(\begin{array}{cc} G_l & 0 \cr \ast & \ast\end{array}\right)=G_{p^\lambda},
\end{equation}
where $G_l$ is a lower triangular matrix over $\mathbb{F}_{p}$ of size $l\times l$. Then we have the following
conclusions}.

\begin{description}
\item{(i)}
 \textit{$G_{p^\lambda}^2=I_{p^\lambda}$}.

\item{(ii)}
 \textit{$G_{l}^2=I_{l}$, ${\rm rank}(G_{l}-I_{l})=\lfloor\frac{l}{2}\rfloor$
and ${\rm rank}(G_{l}+I_{l})=\lceil\frac{l}{2}\rceil$}.

\item{(iii)}
  Denote $x^{-1}=x^{p^\lambda-1}$ (mod $(x-1)^{l}$). For any $B_l=(b_0,b_1,\ldots,b_{l-1})^{{\rm tr}}\in
\mathbb{F}_{p^m}^l$, we denote $b(x)=\sum_{i=0}^{l-1}b_i(x-1)^i$. Then
$$x^{-1}b(x^{-1})\equiv (1,(x-1),\ldots,(x-1)^{l-1})(G_lB_l)
\ ({\rm mod} \ (x-1)^l).$$

\item{(iv)}
  \textit{Let $\Upsilon_1,\Upsilon_2,\ldots,\Upsilon_{l}$ be the column vectors of $G_{l}+I_{l}$, i.e.,
$\Upsilon_i\in \mathbb{F}_{p}^{l}$ for all $i$ and $G_{l}+I_{l}=(\Upsilon_1,\Upsilon_2,\ldots,\Upsilon_{l})$. Then
$\{\Upsilon_{2t-1}\mid t=1,2,\ldots,\lceil\frac{l}{2}\rceil\}$ is an
$\mathbb{F}_{p^m}$-basis of $\mathcal{S}_{l}$. Precisely, we have
${\rm dim}_{\mathbb{F}_{p^m}}(\mathcal{S}_l)=\lceil\frac{l}{2}\rceil$ and}
\begin{equation}
\label{eq5}
\mathcal{S}_{l}=\left\{\sum_{t=1}^{\lceil\frac{l}{2}\rceil}a_{2t-2}\Upsilon_{2t-1}\mid
a_{2t-2}\in \mathbb{F}_{p^m}, \ 1\leq t\leq\lceil\frac{l}{2}\rceil\right\}.
\end{equation}

\end{description}

\noindent
  {\bf Proof.} (i) First, we prove that $G_{p}^2=I_p$ over $\mathbb{F}_p$.
 Since $G_{p}$ is a lower triangular matrix of order $p$,
$G_{p}^2$ is a lower triangular matrix and its
$(i,i)$-entry is
$$\left(g_{i,i}^{(p)}\right)^2=\left((-1)^{i-1}\left(\begin{array}{c}p-i\cr 0\end{array}\right)\right)^2=1,
\ i=1,2,\ldots,p,$$
by Equation (\ref{eq2}) in Section 2. Now, let $1\leq j\leq p-1$ and $i=j+k$ where $1\leq k\leq p-j$. By the
definition of $G_{p}$ and the following combinatorial identities
$$\left(\begin{array}{c}p-j\cr h\end{array}\right)
\left(\begin{array}{c}(p-j)-h\cr k-h\end{array}\right)
=\left(\begin{array}{c}p-j\cr k\end{array}\right)\left(\begin{array}{c}k\cr h\end{array}\right),$$
and $\sum_{h=0}^k(-1)^{h}\left(\begin{array}{c}k\cr h\end{array}\right)=0$, the $(i,j)$-entry of $G_{p}^2$ is equal to
\begin{eqnarray*}
\sum_{z=j}^ig_{i,z}^{(p)}g_{z,j}^{(p)}
 &=&\sum_{h=0}^kg_{j+k,j+h}^{(p)}g_{j+h,j}^{(p)} \\
 &=&\sum_{h=0}^k(-1)^{j+h-1}\left(\begin{array}{c}p-j-h\cr j+k-j-h\end{array}\right)
 \cdot(-1)^{j-1}\left(\begin{array}{c}p-j \cr j+h-j\end{array}\right) \\
 &=&\sum_{h=0}^k(-1)^{h}\left(\begin{array}{c}(p-j)-h\cr k-h\end{array}\right)\left(\begin{array}{c}p-j\cr h\end{array}\right)\\
 &=&\left(\begin{array}{c}p-j\cr k\end{array}\right)\sum_{h=0}^k(-1)^{h}\left(\begin{array}{c}k\cr h\end{array}\right)\\
 &= & 0.
\end{eqnarray*}
As stated above, we conclude that $G_p^2=I_p$. Now, let $\lambda\geq 2$ and assume that $G_{p^{\lambda-1}}^2=I_{p^{\lambda-1}}$
(mod $p$). Then by Propostion 2, it follows that
$$G_{p^{\lambda}}^2=(G_p\otimes G_{p^{\lambda-1}})^2=G_p^2\otimes G_{p^{\lambda-1}}^2=I_p\otimes I_{p^{\lambda-1}}=I_{p^\lambda}.$$

\par
  (ii) By (i) and Equation (\ref{eq4}) for the definition of $G_l$, it follows that
$$\left(\begin{array}{cc} G_l^2 & 0 \cr \ast & \ast\end{array}\right)=G_{p^\lambda}^2=I_{p^\lambda}
=\left(\begin{array}{cc} I_l & 0 \cr \ast & \ast\end{array}\right).$$
This implies $G_l^2=I_l$, and hence $(G_l-I_l)(G_l+I_l)=0$. From this and by linear algebra theory,
we deduce that
$${\rm rank}(G_{l}-I_{l})+{\rm rank}(G_{l}+I_{l})\leq l.$$
Since $G_{p^\lambda}$ is a lower triangular matrix of order $p^\lambda$ with diagonal entries:
$$g_{i,i}^{(p^\lambda)}=(-1)^{i-1}, \ i=1,2,\ldots,p^\lambda,$$
by Equation (\ref{eq4}) we see that
$G_{l}-I_{l}$ is a lower triangular matrix with $\lfloor\frac{l}{2}\rfloor$ nonzero
diagonal entries:
$$g_{2t,2t}^{(p^\lambda)}-1=(-1)^{2t-1}-1=-2, \ t=1,\ldots,\lfloor\frac{l}{2}\rfloor,$$
and
$G_{l}+I_{l}$ is a lower triangular matrix with $\lceil\frac{l}{2}\rceil$ nonzero
diagonal entries:
$$g_{2t-1,2t-1}^{(p^\lambda)}+1=(-1)^{2t-2}+1=2, \ t=1,\ldots,\lceil\frac{l}{2}\rceil.$$
These imply ${\rm rank}(G_{l}-I_{l})\geq \lfloor\frac{l}{2}\rfloor$
and ${\rm rank}(G_{l}+I_{l})\geq \lceil\frac{l}{2}\rceil$.
From these and by $\lfloor\frac{l}{2}\rfloor+\lceil\frac{l}{2}\rceil=l$, we deduce
that ${\rm rank}(G_{l}-I_{l})=\lfloor\frac{l}{2}\rfloor$
and ${\rm rank}(G_{l}+I_{l})=\lceil\frac{l}{2}\rceil$.

\par
  (iii) In the following, we denote
$$X_l=(1,(x-1),(x-1)^2,\ldots,(x-1)^{l-1}),$$
for any integer $l$: $1\leq l\leq p^\lambda$. Then $b(x)=X_lB_l$. By $x^{p^\lambda}\equiv 1$ (mod $(x-1)^{p^\lambda}$)
in $\mathbb{F}_{p^m}[x]$, it follows that
$$
x^{-j}\equiv x^{p^\lambda-j} \ ({\rm mod} \ (x-1)^{p^\lambda}), \ j=0,1,2,\ldots,p^\lambda.
$$
From this and by Equation (\ref{eq2}), we deduce that
\begin{eqnarray*}
x^{-j}(1-x)^{j-1}& \equiv &\left(1+(x-1)\right)^{p^\lambda-j}\cdot(-1)^{j-1}(x-1)^{j-1}\\
 &\equiv &(-1)^{j-1}(x-1)^{j-1}\cdot \sum_{k=0}^{p^\lambda-j}\left(\begin{array}{c}p^\lambda-j\cr k\end{array}\right)(x-1)^k\\
 &\equiv & \sum_{i=j}^{p^\lambda}(-1)^{j-1}\left(\begin{array}{c}p^\lambda-j\cr i-j\end{array}\right)(x-1)^{i-1} \
 ({\rm set} \ i=k+j)\\
 &\equiv & X_{p^\lambda}\cdot  \left(\begin{array}{c}0\cr \ldots \cr 0\cr g^{(p^\lambda)}_{j,j} \cr \ldots \cr g^{(p^\lambda)}_{p^\lambda,j} \end{array}\right) \ ({\rm mod} \ (x-1)^{p^\lambda}),
\end{eqnarray*}
for all $j=1,2,\ldots,p^\lambda$. These imply
\begin{equation}
\label{eq6}
(x^{-1},x^{-2}(1-x),\ldots,x^{-p^\lambda}(1-x)^{p^\lambda-1})\equiv X_{p^\lambda}G_{p^\lambda}
\ ({\rm mod} \ (x-1)^{p^\lambda})
\end{equation}
by the definition of the matrix $G_{p^\lambda}$ in Section 2.

\par
  As $l\leq p^l$, by Equations (\ref{eq6}) and (\ref{eq4}) it follows that
\begin{eqnarray*}
x^{-1}b(x^{-1})
 &=&x^{-1}(1,(x^{-1}-1),(x^{-1}-1)^2,\ldots,(x^{-1}-1)^{l-1})B_l \\
 &=&x^{-1}(1,(x^{-1}-1),(x^{-1}-1)^2,\ldots,(x^{-1}-1)^{p^\lambda-1})\left(\begin{array}{c}B_l \cr 0\end{array}\right) \\
 &=& \left(x^{-1},x^{-2}(1-x),x^{-3}(1-x)^{2},x^{-p^\lambda}(1-x)^{p^\lambda-1}\right)\left(\begin{array}{c}B_l \cr 0\end{array}\right)\\
 &=& X_{p^\lambda}\cdot G_{p^\lambda}\left(\begin{array}{c}B_l \cr 0\end{array}\right) \\
 &\equiv & (X_l,0)\left(\begin{array}{cc}G_l & 0\cr \ast & \ast\end{array}\right)\left(\begin{array}{c}B_l \cr 0\end{array}\right)\\
 &\equiv & X_l(G_lB_l) \ ({\rm mod} \ (x-1)^l).
\end{eqnarray*}

\par
  (iv) Let $\mathcal{H}_l$ be the solution space of following linear equations over $\mathbb{F}_{p^m}$:
$$
(G_l-I_l)Y=0, \ {\rm where} \ Y=(y_0,y_1,\ldots,y_{l-1})^{{\rm tr}}.
$$
By (ii) we have ${\rm dim}_{\mathbb{F}_{p^m}}(\mathcal{H}_{l})=l-{\rm rank}(G_l-I_l)=l-\lfloor\frac{l}{2}\rfloor=\lceil\frac{l}{2}\rceil$.

\par
 By (ii) we know that ${\rm rank}(G_{l}+I_{l})=\lceil\frac{l}{2}\rceil$. This implies
that $\lceil\frac{l}{2}\rceil$ is the rank of the vectors $\Upsilon_1,\Upsilon_2,\ldots,\Upsilon_{l}$.
From this, by
$$\left(\begin{array}{cc} (\Upsilon_1,\Upsilon_2,\ldots,\Upsilon_{l}) & 0 \cr \ast & \ast\ast\end{array}\right)
 =\left(\begin{array}{cc} G_l+I_l & 0 \cr \ast & \ast\ast\end{array}\right)
 = G_{p^\lambda}+I_{p^\lambda}$$
and $G_{p^\lambda}+I_{p^\lambda}=\left(
\begin{array}{cccccc} 2 & & & & & \cr
\ast & 0 & & & & \cr \ast & \ast & 2 & & & \cr \vdots & \vdots & \vdots & \ddots & & \cr
\ast & \ast & \ast & \ldots & 0 & \cr \ast & \ast & \ast &\ldots & \ast & 2\end{array}\right)$ we
deduce that the set of column vectors
$\{\Upsilon_{2t-1}\mid t=1,2,\ldots,\lceil\frac{l}{2}\rceil\}$
 is a
maximal independent system of $\Upsilon_1,\Upsilon_2,\ldots,\Upsilon_{l}$.

\par
  On the other hand, by $(G_{l}-I_{l})(G_{l}+I_{l})=G_l^2-I_l=0$ and
$G_{l}+I_{l}=(\Upsilon_1,\Upsilon_2,\ldots, \Upsilon_{l})$, it follows that
$$(G_{l}-I_{l})\Upsilon_i=0, \ i=1,2,\ldots,l.$$
This implies $\Upsilon_i\in \mathcal{H}_{l}$ for all $i=1,2,\ldots,l$.
Then by ${\rm dim}_{\mathbb{F}_{p^m}}(\mathcal{H}_{l})=\lceil\frac{l}{2}\rceil$, we conclude that
$\{\Upsilon_{2t-1}\mid t=1,2,\ldots,\lceil\frac{l}{2}\rceil\}$ is an $\mathbb{F}_{p^m}$-basis of
$\mathcal{H}_{l}$.

\par
  Now, let's prove that $\mathcal{S}_{l}=\mathcal{H}_{l}$.
Let $b(x)\in \frac{\mathbb{F}_{p^m}[x]}{\langle (x-1)^l\rangle}$ where $1\leq l\leq p^\lambda$. Then $b(x)$ has a unique $(x-1)$-expansion:
$$
b(x)=\sum_{i=0}^{l-1}b_i(x-1)^i=X_lB_l, \ {\rm where} \ B_l=(b_0,b_1,\ldots,b_{l-1})^{{\rm tr}}\in \mathbb{F}_{p^m}^l.
$$
From this and by (iii), we deduce that
$$x^{-1}b(x^{-1})-b(x)=X_l(G_lB_l)-X_lB_l=X_l(G_l-I_l)B_l.$$
Therefore, by Equations (\ref{eq1}) and (\ref{eq3}) we see that
\begin{eqnarray*}
B_l\in \mathcal{S}_l
  &\Leftrightarrow & b(x)\in \Omega_l, \ {\rm i.e.}, \ x^{-1}b(x^{-1})-b(x)\equiv 0 \ ({\rm mod} \ (x-1)^l) \\
  &\Leftrightarrow & (G_l-I_l)B_l=0 \\
  &\Leftrightarrow & B_l\in \mathcal{H}_l.
\end{eqnarray*}
As stated above. we conclude that $\mathcal{S}_l=\mathcal{H}_l$.
\hfill
$\Box$

%%%%%%%%%%%%%%%%%%%%%%%%%%%%%%%%%%%%%%%%%%%%%%%%%%%%%%%%%%%%%%%%%%%%%%%%%%%%%%

%%%%%%%%%%%%%%%%%%%%%%%%%%%%%%%%%%%%%%%%%%%%%%%%%%%%%%%%%%%%%%%%%%%%%%%%%%%%%%%%%%%%%%%%%%

%%%%%%%%%%%%%%%%%%%%%%%%%%%%%%%%%%%%%%%%%%%%%%%%%%%%%%%%%%%%%%%%%%%%%%%%%%%%%%%%%%%%%%%%%%%%%%
\section{Construction and enumeration for self-dual cyclic codes} \label{}

\noindent
In this section, we determine all self-dual cyclic codes of length $p^s$ over $\mathbb{F}_{p^m}+u\mathbb{F}_{p^m}$ ($u^2=0$).

\vskip 3mm\noindent
 {\bf Theorem 2}
   \textit{For any integers $l$ and $\delta$, $0\leq \delta<l\leq p^s-1$, we denote
\begin{equation}
\label{eq7}
\mathcal{S}_l^{[\delta]}=\{B_l^{[\delta]}=(b_\delta, \ldots, b_{l-1})^{{\rm tr}}\mid
(0,\ldots,0,b_\delta, \ldots, b_{l-1})^{{\rm tr}} \in\mathcal{S}_l\},
\end{equation}
and write}
$G_{l}+I_{l}=(\Upsilon_1,\Upsilon_2,\ldots, \Upsilon_{l})$,
\textit{where $\Upsilon_j\in \mathbb{F}_{p^m}^l$ for all $j=1,2,\ldots,l$. Moreover, for any integer $j$,
$\lceil \frac{\delta}{2}\rceil+1\leq j\leq \lceil \frac{l}{2}\rceil$, we set
$$\Upsilon_{2j-1}^{[\delta;l)}=\left(\begin{array}{c}g_{\delta+1,2j-1}
\cr g_{\delta+2,2j-1}\cr\ldots \cr g_{l,2j-1}\end{array}\right),
\ {\rm when} \ \Upsilon_{2j-1}=\left(\begin{array}{c} g_{1,2j-1} \cr \ldots\cr g_{\delta,2j-1}\cr\hline
g_{\delta+1,2j-1}
\cr g_{\delta+2,2j-1}
\cr\ldots \cr g_{l,2j-1}\end{array}\right).$$
Then}
$$\mathcal{S}_l^{[\delta]}=\left\{\sum_{\lceil\frac{\delta}{2}\rceil+1\leq j\leq\lceil\frac{l}{2}\rceil}a_{2j-2}\Upsilon_{2j-1}^{[\delta;l)}
\mid a_{2j-2}\in \mathbb{F}_{p^m}, \ \lceil\frac{\delta}{2}\rceil+1\leq j\leq\lceil\frac{l}{2}\rceil\right\}.$$
\textit{Hence $|\mathcal{S}_l^{[\delta]}|=p^{(\lceil\frac{l}{2}\rceil-\lceil\frac{\delta}{2}\rceil)m}$}.

\vskip 3mm\noindent
  {\bf Proof.} By Equation (\ref{eq4}) and the proof of Theorem 1(iv), we have
\begin{equation}
\label{eq8}
\Upsilon_1=\left(\begin{array}{c}2\cr g_{2,1}\cr  g_{2,1}\cr \vdots \cr g_{l,1}\end{array}\right),
\ \Upsilon_{2j-1}=\left(\begin{array}{c}\textbf{0}_{(2j-2)\times 1}\cr 2\cr g_{2j,2j-1}\cr \vdots \cr g_{l,2j-1}\end{array}\right)
\ {\rm for} \ {\rm all} \ 2\leq j\leq \lceil\frac{l}{2}\rceil,
\end{equation}
where $\textbf{0}_{(2j-2)\times 1}$ is the zero matrix of type $(2j-2)\times 1$ and $g_{i,j}\in\mathbb{F}_p$.

\par
  Let $B_l=(b_0,b_1,\ldots,b_{l-1})^{{\rm tr}}\in \mathbb{F}_{p^m}^l$. Since $\{\Upsilon_{2t-1}\mid t=1,2,\ldots,\lceil\frac{l}{2}\rceil\}$ is an
$\mathbb{F}_{p^m}$-basis of $\mathcal{S}_{l}$ by Theorem 1(iv), we see that
$B_l\in\mathcal{S}_l$ if and only if there is a unique
row vector $(a_0,a_2,\ldots,a_{2\lceil\frac{l}{2}\rceil-2})$ over $\mathbb{F}_{p^m}$ such that
\begin{eqnarray*}
B_l &=& \sum_{j=1}^{\lceil\frac{l}{2}\rceil}a_{2j-2}\Upsilon_{2j-1}
 =\left(\begin{array}{c}2a_0\cr g_{2,1}a_0\cr  g_{2,1}a_0\cr \vdots \cr g_{l,1}a_0\end{array}\right)
  +\sum_{2\leq j\leq \lceil\frac{l}{2}\rceil}
  \left(\begin{array}{c}\textbf{0}_{(2j-2)\times 1}\cr 2a_{2j-2}\cr g_{2j,2j-1}a_{2j-2}\cr \vdots \cr g_{l,2j-1}a_{2j-2}\end{array}\right)\\
 &=&\left(\begin{array}{l}2a_0\cr g_{2,1}a_0\cr g_{3,1}a_0+2a_2 \cr g_{4,1}a_0+g_{4,3}a_2\cr\vdots
\cr g_{2j-1,1}a_0+g_{2j-1,3}a_2+\ldots+g_{2j-1,2j-3}a_{2j-4}+2a_{2j-2}
\cr g_{2j,1}a_0+g_{2j,3}a_2+\ldots+g_{2j,2j-3}a_{2j-4}+g_{2j,2j-1}a_{2j-2}\cr \vdots
\cr \sum_{j=1}^{\lceil\frac{l}{2}\rceil}g_{l,2j-1}a_{2j-2}\end{array}\right).
\end{eqnarray*}
From this we deduce that
\begin{eqnarray*}
&& B_l^{[\delta]}=(b_\delta, \ldots, b_{l-1})^{{\rm tr}}\in \mathcal{S}_l^{[\delta]} \\
&\Leftrightarrow & B_l=(b_0,b_1,\ldots,b_{l-1})^{{\rm tr}}\in\mathcal{S}_l \ {\rm satisfying} \ b_0=b_1=\ldots=b_{\delta-1}=0\\
&\Leftrightarrow & B_l=\sum_{\lceil\frac{\delta}{2}\rceil+1\leq j\leq\lceil\frac{l}{2}\rceil}a_{2j-2}\Upsilon_{2j-1}\\
&\Leftrightarrow & B_l^{[\delta]}
=\sum_{\lceil\frac{\delta}{2}\rceil+1\leq j\leq\lceil\frac{l}{2}\rceil}a_{2j-2}\Upsilon_{2j-1}^{[\delta;l)}.
\end{eqnarray*}
This implies
$$\mathcal{S}_l^{[\delta]}=\left\{\sum_{\lceil\frac{\delta}{2}\rceil+1\leq j\leq\lceil\frac{l}{2}\rceil}a_{2j-2}\Upsilon_{2j-1}^{[\delta;l)}
\mid a_{2j-2}\in \mathbb{F}_{p^m}, \ \lceil\frac{\delta}{2}\rceil+1\leq j\leq\lceil\frac{l}{2}\rceil\right\}.$$

\par
  Finally, by Equation (\ref{eq8}) and
$$2(\lceil \frac{\delta}{2}\rceil+1)-1=2\lceil \frac{\delta}{2}\rceil+1
=\left\{\begin{array}{ll} \delta+1, & {\rm if} \ \delta \ {\rm is} \ {\rm even}; \cr
\delta+2, & {\rm if} \ \delta \ {\rm is} \ {\rm odd},\end{array}\right.$$
we conclude that $\{\Upsilon_{2j-1}^{[\delta;l)}\mid \lceil\frac{\delta}{2}\rceil+1\leq j\leq\lceil\frac{l}{2}\rceil\}$
is a linear independent set of vectors in $\mathbb{F}_{p^m}^{l-\delta}$.
Therefore, ${\rm dim}_{\mathbb{F}_{p^m}}(\mathcal{S}_l^{[\delta]})=\lceil \frac{l}{2}\rceil-\lceil \frac{\delta}{2}\rceil$ and hence
$|\mathcal{S}_l^{[\delta]}|=p^{(\lceil \frac{l}{2}\rceil-\lceil \frac{\delta}{2}\rceil)m}$.
\hfill
$\Box$

\vskip 3mm\par
  Now, we list explicitly all self-dual cyclic codes of length $p^s$ over $\mathbb{F}_{p^m}+u\mathbb{F}_{p^m}$ ($u^2=0$)
by the following theorem.

\vskip 3mm\noindent
 {\bf Theorem 3}
   \textit{Using the notation in Theorem 2, let $p$ be any odd prime number and $s$ be any positive integer. Then
we have the following conclusions}.

\begin{description}
\item{($\dag$)}
  \textit{If $p^s\equiv 3$ (mod $4$), all
distinct self-dual cyclic codes of length $p^s$ over $\mathbb{F}_{p^m}+u\mathbb{F}_{p^m}$ are given by the following two cases}.

$\bullet$ \textit{$\sum_{\nu=0}^{\frac{p^s+1}{4}-1}(p^m)^{\frac{p^s+1}{4}-1-\nu}$ codes:
$$\langle (x-1)^{2\nu+1}b(x)+u(x-1)^{2\nu}, (x-1)^{p^s-2\nu}\rangle,$$
where $b(x)=\sum_{i=\frac{p^s-1}{2}-2\nu}^{p^s-1-4\nu-1}b_i(x-1)^i$ with}
$$\left(\begin{array}{c}b_{\frac{p^s-1}{2}-2\nu} \cr
b_{\frac{p^s-1}{2}-2\nu+1}\cr \ldots \cr
b_{p^s-1-4\nu-1}\end{array}\right)=\sum_{\frac{p^s+1}{4}-\nu+1\leq j\leq \frac{p^s+1}{2}-2\nu-1}a_{2j-2}
\Upsilon_{2j-1}^{[\frac{p^s-1}{2}-2\nu; p^s-1-4\nu)}$$
\textit{for $a_{2j-2}\in \mathbb{F}_{p^m}$ arbitrary and $0\leq \nu\leq \frac{p^s+1}{4}-1$}.

$\bullet$ \textit{$\sum_{\nu=0}^{\frac{p^s+1}{4}-1}(p^m)^{\frac{p^s+1}{4}-1-\nu}$ codes:
$$\langle (x-1)^{2\nu+2}b(x)+u(x-1)^{2\nu+1}, (x-1)^{p^s-2\nu-1}\rangle,$$
where $b(x)=\sum_{i=\frac{p^s-1}{2}-2\nu-1}^{p^s-1-4\nu-3}b_i(x-1)^i$ with}
$$\left(\begin{array}{c}b_{\frac{p^s-1}{2}-2\nu-1} \cr
b_{\frac{p^s-1}{2}-2\nu}\cr \ldots \cr
b_{p^s-1-4\nu-3}\end{array}\right)=\sum_{\frac{p^s+1}{4}-\nu\leq j\leq \frac{p^s+1}{2}-2\nu-2}a_{2j-2}
\Upsilon_{2j-1}^{[\frac{p^s-1}{2}-2\nu-1; p^s-1-4\nu-2)}$$
\textit{for $a_{2j-2}\in \mathbb{F}_{p^m}$ arbitrary and $0\leq \nu\leq \frac{p^s+1}{4}-1$}.

\textit{In this case, the number of all self-dual cyclic codes of length $p^s$ over $\mathbb{F}_{p^m}+u\mathbb{F}_{p^m}$
is equal to}
$${\rm NE}(\mathbb{F}_{p^m}+u\mathbb{F}_{p^m},p^s)=2\left(\frac{(p^m)^{\frac{p^s+1}{4}}-1}{p^m-1}\right).$$

\item{($\ddag$)}
  \textit{If $p^s\equiv 1$ (mod $4$), all
distinct self-dual cyclic codes of length $p^s$ over $\mathbb{F}_{p^m}+u\mathbb{F}_{p^m}$ are given by the following three cases}:

$\bullet$ \textit{$(p^m)^{\frac{p^s-1}{4}}$ codes}:
$$\langle (x-1)b(x)+u\rangle,$$
\textit{where
$b(x)=\sum_{i=\frac{p^s-1}{2}}^{p^s-2}b_i(x-1)^i$ with}
$$\left(\begin{array}{c}b_{\frac{p^s-1}{2}} \cr
b_{\frac{p^s-1}{2}+1}\cr \ldots \cr
b_{p^s-2}\end{array}\right)=\sum_{\frac{p^s-1}{4}+1\leq j\leq \frac{p^s-1}{2}}a_{2j-2}
\Upsilon_{2j-1}^{[\frac{p^s-1}{2}; p^s-1)}$$
\textit{for $a_{2j-2}\in \mathbb{F}_{p^m}$ arbitrary}.

$\bullet$ \textit{$\sum_{\nu=1}^{\frac{p^s-1}{4}}(p^m)^{\frac{p^s-1}{4}-\nu}$ codes:
$$\langle (x-1)^{2\nu+1}b(x)+u(x-1)^{2\nu}, (x-1)^{p^s-2\nu}\rangle,$$
where $b(x)=\sum_{i=\frac{p^s-1}{2}-2\nu}^{p^s-1-4\nu-1}b_i(x-1)^i$ with}
$$\left(\begin{array}{c}b_{\frac{p^s-1}{2}-2\nu} \cr
b_{\frac{p^s-1}{2}-2\nu+1}\cr \ldots \cr
b_{p^s-1-4\nu-1}\end{array}\right)=\sum_{\frac{p^s-1}{4}-\nu+1\leq j\leq \frac{p^s-1}{2}-2\nu}a_{2j-2}
\Upsilon_{2j-1}^{[\frac{p^s-1}{2}-2\nu; p^s-1-4\nu)}$$
\textit{for $a_{2j-2}\in \mathbb{F}_{p^m}$ arbitrary and $1\leq \nu\leq \frac{p^s-1}{4}$}.

$\bullet$ \textit{$\sum_{\nu=1}^{\frac{p^s-1}{4}}(p^m)^{\frac{p^s-1}{4}-\nu}$ codes:
$$\langle (x-1)^{2\nu}b(x)+u(x-1)^{2\nu-1}, (x-1)^{p^s-2\nu+1}\rangle,$$
where $b(x)=\sum_{i=\frac{p^s-1}{2}-2\nu+1}^{p^s-4\nu}b_i(x-1)^i$ with}
$$\left(\begin{array}{c}b_{\frac{p^s-1}{2}-2\nu+1} \cr
b_{\frac{p^s-1}{2}-2\nu+2}\cr \ldots \cr
b_{p^s-4\nu}\end{array}\right)=\sum_{\frac{p^s-1}{4}-\nu+2\leq j\leq \frac{p^s-1}{2}-2\nu+1}a_{2j-2}
\Upsilon_{2j-1}^{[\frac{p^s-1}{2}-2\nu+1; p^s+1-4\nu)}$$
\textit{for $a_{2j-2}\in \mathbb{F}_{p^m}$ arbitrary and $1\leq \nu\leq \frac{p^s-1}{4}$}.

\textit{In this case, the number of all self-dual cyclic codes of length $p^s$ over $\mathbb{F}_{p^m}+u\mathbb{F}_{p^m}$
is equal to}
$${\rm NE}(\mathbb{F}_{p^m}+u\mathbb{F}_{p^m},p^s)=(p^m)^{\frac{p^s-1}{4}}+2\left(\frac{(p^m)^{\frac{p^s-1}{4}}-1}{p^m-1}\right).$$
\end{description}

\noindent
  {\bf Proof.}
  By Proposition 1, all distinct
self-dual codes of length $p^s$ over $\mathbb{F}_{p^m}+u\mathbb{F}_{p^m}$ are given as follows:
$$\langle (x-1)^{k+1}b(x)+u(x-1)^k, (x-1)^{p^s-k}\rangle,$$
where
$0\leq k\leq \frac{p^s-1}{2}$ and  $b(x)=\sum_{i=\frac{p^s-1}{2}-k}^{p^s-2-2k}b_i(x-1)^i\in \Omega_{p^s-1-2k}$.

\par
  Let $b(x)=\sum_{i=\frac{p^s-1}{2}-k}^{p^s-2-2k}b_i(x-1)^i$ where
$b_i\in \mathbb{F}_{p^m}$ for all $i=\frac{p^s-1}{2}-k,\ldots, p^s-2-2k$. Then by Equations (\ref{eq3})
and (\ref{eq7}), it follows that
\begin{eqnarray*}
b(x)\in \Omega_{p^s-1-2k} &\Leftrightarrow& (0,\ldots,0, b_{\frac{p^s-1}{2}-k},\ldots,b_{p^s-2-2k})\in \mathcal{S}_{p^s-1-2k}\\
 &\Leftrightarrow& (b_{\frac{p^s-1}{2}-k},\ldots,b_{p^s-2-2k})\in \mathcal{S}_{p^s-1-2k}^{[\frac{p^s-1}{2}-k]}.
\end{eqnarray*}
Moreover, by Theorem 2 we have
$|\mathcal{S}_{p^s-1-2k}^{[\frac{p^s-1}{2}-k]}|=(p^m)^{\lceil \frac{p^s-1-2k}{2}\rceil
-\lceil\frac{\frac{p^s-1}{2}-k}{2}\rceil}$.
Hence the number of all self-dual cyclic codes of length $p^s$ over $\mathbb{F}_{p^m}+u\mathbb{F}_{p^m}$
is equal to ${\rm NE}(\mathbb{F}_{p^m}+u\mathbb{F}_{p^m},p^s)=\sum_{k=0}^{\frac{p^s-1}{2}}|\mathcal{S}_{p^s-1-2k}^{[\frac{p^s-1}{2}-k]}|$.

\par
  Now, we have the following two situations.

\par
  ($\dag$) Let $p^s\equiv 3$ (mod $4$). Then $\frac{p^s-1}{2}$ is odd and $\frac{\frac{p^s-1}{2}+1}{2}=\frac{p^s+1}{4}$ is a positive integer. In this situation,  we have one of the following two cases:

\par
  ($\dag$-i) Let $k=2\nu$ be even, where $0\leq \nu\leq \frac{p^s+1}{4}-1$. Then
$$\left\lceil\frac{\frac{p^s-1}{2}-k}{2}\right\rceil=\left\lceil\frac{\frac{p^s-3}{2}+1-2\nu}{2}\right\rceil
=\frac{p^s-3}{4}+1-\nu$$
and $\lceil \frac{p^s-1-4\nu}{2}\rceil
-\lceil\frac{\frac{p^s-1}{2}-2\nu}{2}\rceil=\frac{p^s-3}{2}+1-2\nu
-(\frac{p^s-3}{4}+1-\nu)=\frac{p^s-3}{4}-\nu$.
From these, we deduce the following conclusions:
\begin{description}
\item{$\diamond$}
  $\mathcal{S}_{p^s-1-2k}^{[\frac{p^s-1}{2}-k]}=\mathcal{S}_{p^s-1-4\nu}^{[\frac{p^s-1}{2}-2\nu]}$.

\item{$\diamond$}
 $|\mathcal{S}_{p^s-1-4\nu}^{[\frac{p^s-1}{2}-2\nu]}|=(p^m)^{\lceil \frac{p^s-1-4u}{2}\rceil
-\lceil\frac{\frac{p^s-1}{2}-2\nu}{2}\rceil}=(p^m)^{\frac{p^s-3}{4}-\nu}=(p^m)^{\frac{p^s+1}{4}-1-\nu}$.
\end{description}

\noindent
  Moreover, by Theorem 2 we have
\begin{eqnarray*}
\mathcal{S}_{p^s-1-4\nu}^{[\frac{p^s-1}{2}-2\nu]}
 &=&\left\{\sum_{\frac{p^s+1}{4}-\nu+1\leq j\leq \frac{p^s+1}{2}-2\nu-1}a_{2j-2}
\Upsilon_{2j-1}^{[\frac{p^s-1}{2}-2\nu; p^s-1-4\nu)} \right.\\
 && \ \ \left.\mid a_{2j-2}\in \mathbb{F}_{p^m},
\ \frac{p^s+1}{4}-\nu+1\leq j\leq \frac{p^s+1}{2}-2\nu-1\right\}.
\end{eqnarray*}

\par
  ($\dag$-ii) Let $k=2\nu+1$ be odd, where $0\leq \nu\leq \frac{p^s+1}{4}-1$. Then
$$\left\lceil\frac{\frac{p^s-1}{2}-k}{2}\right\rceil=\left\lceil\frac{\frac{p^s-3}{2}+1-2\nu-1}{2}\right\rceil
=\frac{p^s-3}{4}-\nu$$
and $\lceil \frac{p^s-1-4u-2}{2}\rceil
-\lceil\frac{\frac{p^s-1}{2}-2\nu-1}{2}\rceil=\frac{p^s-3}{2}+1-2\nu-1
-(\frac{p^s-3}{4}-\nu)=\frac{p^s-3}{4}-\nu$
From these, we deduce the following conclusions:
\begin{description}
\item{$\diamond$}
  $\mathcal{S}_{p^s-1-2k}^{[\frac{p^s-1}{2}-k]}=\mathcal{S}_{p^s-1-4\nu-2}^{[\frac{p^s-1}{2}-2\nu-1]}$.

\item{$\diamond$}
 $|\mathcal{S}_{p^s-1-4\nu-2}^{[\frac{p^s-1}{2}-2\nu-1]}|=(p^m)^{\lceil \frac{p^s-1-4u-2}{2}\rceil
-\lceil\frac{\frac{p^s-1}{2}-2\nu-1}{2}\rceil}=(p^m)^{\frac{p^s+1}{4}-1-\nu}$.
\end{description}
\noindent
  Moreover, by Theorem 2 we have
\begin{eqnarray*}
\mathcal{S}_{p^s-1-4\nu-2}^{[\frac{p^s-1}{2}-2\nu-1]}
 &=&\left\{\sum_{\frac{p^s+1}{4}-\nu\leq j\leq \frac{p^s+1}{2}-2\nu-2}a_{2j-2}
\Upsilon_{2j-1}^{[\frac{p^s-1}{2}-2\nu-1; p^s-1-4\nu-2)} \right.\\
 && \ \ \left.\mid a_{2j-2}\in \mathbb{F}_{p^m},
\ \frac{p^s+1}{4}-\nu\leq j\leq \frac{p^s+1}{2}-2\nu-2\right\}.
\end{eqnarray*}

\par
  Therefore, when $p^s\equiv 3$ (mod $4$) we have
$${\rm NE}(\mathbb{F}_{p^m}+u\mathbb{F}_{p^m},p^s)=2\sum_{\nu=0}^{\frac{p^s+1}{4}-1}(p^m)^{\frac{p^s+1}{4}-1-\nu}
=2\left(\frac{(p^m)^{\frac{p^s+1}{4}}-1}{p^m-1}\right).$$

\par
  ($\ddag$) Let $p^s\equiv 1$ (mod $4$). Then $\frac{p^s-1}{2}$ is even and $\frac{p^s-1}{4}$ is a positive integer. In this situation,  we have one of the following three cases:

\par
  ($\ddag$-i) Let $k=0$. Then we have that
  $$\mathcal{S}_{p^s-1-2k}^{[\frac{p^s-1}{2}-k]}=\mathcal{S}_{p^s-1}^{[\frac{p^s-1}{2}]}
\ {\rm and} \ |\mathcal{S}_{p^s-1}^{[\frac{p^s-1}{2}]}|=(p^m)^{\lceil\frac{p^s-1}{2}\rceil
-\lceil \frac{\frac{p^s-1}{2}}{2}\rceil}=(p^m)^{\frac{p^s-1}{4}}.$$
\noindent
  Moreover, by Theorem 2 we have
\begin{eqnarray*}
\mathcal{S}_{p^s-1}^{[\frac{p^s-1}{2}]}
 &=&\left\{\sum_{\frac{p^s-1}{4}+1\leq j\leq \frac{p^s-1}{2}}a_{2j-2}
\Upsilon_{2j-1}^{[\frac{p^s-1}{2}; p^s-1)} \mid a_{2j-2}\in \mathbb{F}_{p^m}, \right. \\
 && \ \ \left.
\ \frac{p^s-1}{4}+1\leq j\leq \frac{p^s-1}{2}\right\}.
\end{eqnarray*}

\par
  ($\ddag$-ii) Let $k=2\nu$ be even, where $1\leq \nu\leq \frac{p^s-1}{4}$. Then
$$\left\lceil\frac{\frac{p^s-1}{2}-k}{2}\right\rceil=\left\lceil\frac{\frac{p^s-1}{2}-2\nu}{2}\right\rceil
=\frac{p^s-1}{4}-\nu$$
and $\lceil \frac{p^s-1-4u}{2}\rceil
-\lceil\frac{\frac{p^s-1}{2}-2\nu}{2}\rceil=\frac{p^s-1}{2}-2\nu
-(\frac{p^s-1}{4}-\nu)=\frac{p^s-1}{4}-\nu$
From these, we deduce the following conclusions:
\begin{description}
\item{$\diamond$}
  $\mathcal{S}_{p^s-1-2k}^{[\frac{p^s-1}{2}-k]}=\mathcal{S}_{p^s-1-4\nu}^{[\frac{p^s-1}{2}-2\nu]}$.

\item{$\diamond$}
 $|\mathcal{S}_{p^s-1-4\nu}^{[\frac{p^s-1}{2}-2\nu]}|=(p^m)^{\lceil \frac{p^s-1-4u}{2}\rceil
-\lceil\frac{\frac{p^s-1}{2}-2\nu}{2}\rceil}=(p^m)^{\frac{p^s-1}{4}-\nu}$.
\end{description}
\noindent
  Moreover, by Theorem 2 we have
\begin{eqnarray*}
\mathcal{S}_{p^s-1-4\nu}^{[\frac{p^s-1}{2}-2\nu]}
 &=&\left\{\sum_{\frac{p^s-1}{4}-\nu+1\leq j\leq \frac{p^s-1}{2}-2\nu}a_{2j-2}
\Upsilon_{2j-1}^{[\frac{p^s-1}{2}-2\nu; p^s-1-4\nu)} \mid a_{2j-2}\in \mathbb{F}_{p^m}, \right.\\
 && \ \ \left.
\ \frac{p^s-1}{4}-\nu+1\leq j\leq \frac{p^s-1}{2}-2\nu\right\}.
\end{eqnarray*}

\par
  ($\ddag$-iii) Let $k=2\nu-1$ be odd, where $1\leq \nu\leq \frac{p^s-1}{4}$. Then
$$\left\lceil\frac{\frac{p^s-1}{2}-k}{2}\right\rceil=\left\lceil\frac{\frac{p^s-1}{2}-2\nu+1}{2}\right\rceil
=\frac{p^s-1}{4}-\nu+1$$
and $\lceil \frac{p^s-1-4u+2}{2}\rceil
-\lceil\frac{\frac{p^s-1}{2}-2\nu+1}{2}\rceil=\frac{p^s-1}{2}-2\nu+1
-(\frac{p^s-1}{4}-\nu+1)=\frac{p^s-1}{4}-\nu$
From these, we deduce the following conclusions:
\begin{description}
\item{$\diamond$}
  $\mathcal{S}_{p^s-1-2k}^{[\frac{p^s-1}{2}-k]}=\mathcal{S}_{p^s-1-4\nu+2}^{[\frac{p^s-1}{2}-2\nu+1]}$.

\item{$\diamond$}
 $|\mathcal{S}_{p^s-1-4\nu+2}^{[\frac{p^s-1}{2}-2\nu+1]}|=(p^m)^{\lceil \frac{p^s-1-4u+2}{2}\rceil
-\lceil\frac{\frac{p^s-1}{2}-2\nu+1}{2}\rceil}=(p^m)^{\frac{p^s-1}{4}-\nu}$.
\end{description}
\noindent
  Moreover, by Theorem 2 we have
\begin{eqnarray*}
\mathcal{S}_{p^s-1-4\nu+2}^{[\frac{p^s-1}{2}-2\nu+1]}
 &=&\left\{\sum_{\frac{p^s-1}{4}-\nu+2\leq j\leq \frac{p^s-1}{2}-2\nu+1}a_{2j-2}
\Upsilon_{2j-1}^{[\frac{p^s-1}{2}-2\nu+1; p^s+1-4\nu)} \right.\\
 && \ \ \left.  \mid a_{2j-2}\in \mathbb{F}_{p^m},
\ \frac{p^s-1}{4}-\nu+2\leq j\leq \frac{p^s-1}{2}-2\nu+1\right\}.
\end{eqnarray*}

\par
  Therefore, when $p^s\equiv 1$ (mod $4$) we have
$${\rm NE}(\mathbb{F}_{p^m}+u\mathbb{F}_{p^m},p^s)
=(p^m)^{\frac{p^s-1}{4}}+2\sum_{\nu=1}^{\frac{p^s-1}{4}}(p^m)^{\frac{p^s-1}{4}-\nu},$$
i.e., ${\rm NE}(\mathbb{F}_{p^m}+u\mathbb{F}_{p^m},p^s)
=(p^m)^{\frac{p^s-1}{4}}+2\left(\frac{(p^m)^{\frac{p^s-1}{4}}-1}{p^m-1}\right)$.
\hfill $\Box$

\vskip 3mm
\noindent
  {\bf Remark} Let $\Gamma\in\{\frac{(\mathbb{F}_{p^m}+u\mathbb{F}_{p^m})[x]}{\langle x^{p^s}-1\rangle},
\frac{(\mathbb{F}_{p^m}+u\mathbb{F}_{p^m})[x]}{\langle x^{p^s}+1\rangle}\}$ and define a map
$\tau: \Gamma\rightarrow \Gamma$ by:
$$\tau(\alpha(x))=\alpha(x^{-1}), \ \forall \alpha(x)\in \Gamma,$$
where $x^{-1}=x^{p^s-1}$ in $\frac{(\mathbb{F}_{p^m}+u\mathbb{F}_{p^m})[x]}{\langle x^{p^s}-1\rangle}$
and $x^{-1}=-x^{p^s-1}$ in $\frac{(\mathbb{F}_{p^m}+u\mathbb{F}_{p^m})[x]}{\langle x^{p^s}+1\rangle}$. Then
$\tau$ is a ring automorphism on $\Gamma$. As $p$ is odd, the map $\varphi$ defined by
$$\varphi(\alpha(x))=\alpha(-x) \ (\forall \alpha(x)\in (\mathbb{F}_{p^m}+u\mathbb{F}_{p^m})[x]/\langle x^{p^s}-1\rangle)$$
is an isomorphism of rings from $\frac{(\mathbb{F}_{p^m}+u\mathbb{F}_{p^m})[x]}{\langle x^{p^s}-1\rangle}$
onto $\frac{(\mathbb{F}_{p^m}+u\mathbb{F}_{p^m})[x]}{\langle x^{p^s}+1\rangle}$ such that the diagram
$
\begin{array}{ccc} \ \ \ \ \frac{(\mathbb{F}_{p^m}+u\mathbb{F}_{p^m})[x]}{\langle x^{p^s}-1\rangle}
 & \stackrel{\varphi}{\longrightarrow} &  \frac{(\mathbb{F}_{p^m}+u\mathbb{F}_{p^m})[x]}{\langle x^{p^s}+1\rangle} \cr
  \tau  \downarrow &  & \ \ \ \downarrow \tau \cr
 \ \ \ \ \frac{(\mathbb{F}_{p^m}+u\mathbb{F}_{p^m})[x]}{\langle x^{p^s}-1\rangle} & \stackrel{\varphi}{\longrightarrow} &  \frac{(\mathbb{F}_{p^m}+u\mathbb{F}_{p^m})[x]}{\langle x^{p^s}+1\rangle}
\end{array}$
commutes.

\par
  For any ideal $\mathcal{C}$ of the ring $\Gamma$, its \textit{annihilating ideal} is defined
as
$${\rm Ann}(\mathcal{C})=\{\alpha(x)\in \Gamma\mid \alpha(x)c(x)=0
\ {\rm in} \ \Gamma, \ \forall c(x)\in \mathcal{C}\}.$$
Then it is well known that $\mathcal{C}^{\bot}=\tau({\rm Ann}(\mathcal{C}))$ (cf. \cite{s10}, \cite{s11}), and hence $\mathcal{C}^{\bot}=\mathcal{C}$ if and only if
$\tau({\rm Ann}(\mathcal{C}))=\mathcal{C}$. From these
we deduce that all distinct self-dual negacyclic codes of length $p^s$ over $\mathbb{F}_{p^m}+u\mathbb{F}_{p^m}$ are the following:
\begin{description}
\item{$\bullet$}
 $\varphi(\mathcal{C})=\{\alpha(-x)\mid \alpha(x)\in \mathcal{C}\}$,
 where $\mathcal{C}$ is an arbitrary self-dual cyclic code of length $p^s$
over $\mathbb{F}_{p^m}+u\mathbb{F}_{p^m}$.
\end{description}
Then by Theorem 3, one can list all distinct self-dual negacyclic codes of length $p^s$ over $\mathbb{F}_{p^m}+u\mathbb{F}_{p^m}$
explicitly. Here we omitted.

\section{Self-dual cyclic codes over $\mathbb{F}_{3^m}+u\mathbb{F}_{3^m}$ of length $3^s$ for $s=1,2,3$}
\noindent
  In this section, we show how to list explicitly all distinct
self-dual cyclic codes over $\mathbb{F}_{p^m}+u\mathbb{F}_{p^m}$ of length $p^s$ for some specific odd
prime number $p$ and positive integers $m,s$. Here, we take cases of $p=3$ and $s=1,2,3$ as examples.

\par
  {\bf (I)} Let $s=1$. Then $\frac{3+1}{4}=1$ and $\nu=0$. In this case,
${\rm NE}(\mathbb{F}_{3^m}+u\mathbb{F}_{3^m},3)=2(\frac{3^m-1}{3^m-1})=2$,
and there is no integer $j$
satisfying one of the following two inequalities:

\par
 $2=\frac{3+1}{4}-\nu+1\leq j\leq \frac{3+1}{2}-2\nu-1=1$;
 $1=\frac{3+1}{4}-\nu\leq j\leq \frac{3+1}{2}-2\nu-2=0.$

\noindent
By Theorem 3($\dag$), all self-dual cyclic codes over $\mathbb{F}_{3^m}+u\mathbb{F}_{3^m}$ of length $3$ are given by:
$$\langle u\rangle \ {\rm and} \ \langle u(x-1),(x-1)^2\rangle.$$

\par
  {\bf (II)} Let $s=2$. Then $3^2\equiv 1$ (mod 4) and $\frac{3^2-1}{4}=2$. In this case, the number
of self-dual cyclic codes over $\mathbb{F}_{3^m}+u\mathbb{F}_{3^m}$ of length $3^2$ is
$${\rm NE}(\mathbb{F}_{3^m}+u\mathbb{F}_{3^m},3^2)=(3^m)^2+2(\frac{(3^m)^2-1}{3^m-1})=3^{2m}+2\cdot3^m+2.$$
by Theorem 3($\ddag$). It is clear that
$G_3+I_3=\left(\begin{array}{ccc} 2 & 0 & 0\cr -1 & 0 & 0 \cr 1 & -1 & 2\end{array}\right)$
and
$$G_8+I_8
=\left(\begin{array}{ccc|ccc|cc} 2 & 0 & 0 & 0 & 0 & 0 & 0 & 0
\cr -1 & 0 & 0 & 0 & 0 & 0 & 0 & 0
\cr 1 & -1 & \textbf{2} & 0 & 0 & 0 & 0 & 0 \cr\hline
-1 & 0 & \textbf{0} & 0 & 0 & 0 & 0 & 0
\cr 1 & 1 & 0 & 1 & \textbf{2} & 0 & \textbf{0} & 0  \cr -1 & 1 & -1 & -1 & \textbf{1} & 0 & \textbf{0} & 0
\cr\hline
1 & 0 & 0 & -1 & \textbf{0} & 0 & \textbf{2} & 0
\cr -1 & -1 & 0 & 1 & \textbf{1} & 0 & \textbf{-1} & 0 \end{array}\right).$$

\par
 By Theorem 3($\ddag$), we have the following three cases:
\begin{description}
\item{\textsl{Case 1.}}
   $k=0$. In this case, $3^2-1=8$, $\frac{3^2-1}{2}=4$ and $3=\frac{3^2-1}{4}+1\leq j\leq \frac{3^2-1}{2}=4$. Using the notation
in Theorem 2, we have
$$\Upsilon_5^{[4,8)}=\Upsilon_{2\cdot 3-1}^{[4,8)}=\left(\begin{array}{c}  2 \cr 1 \cr 0 \cr 1\end{array}\right),
\ \Upsilon_7^{[4,8)}=\Upsilon_{2\cdot 4-1}^{[4,8)}=\left(\begin{array}{c}  0 \cr 0 \cr 2 \cr -1\end{array}\right),$$
and
$(b_4,b_5,b_6,b_7)=\left(a_4\Upsilon_5^{[4,8)}+a_6\Upsilon_7^{[4,8)}\right)^{{\rm tr}}
=(2a_4,a_4,2a_6,a_4+2a_6)$. Hence
there are $(3^m)^2$ self-dual cyclic codes over $\mathbb{F}_{3^m}+u\mathbb{F}_{3^m}$ of length $3^2$:

\par
  $\bullet$ $\langle (x-1)b(x)+u\rangle$, where
$$b(x)=2a_4(x-1)^4+a_4(x-1)^5+2a_6(x-1)^6+(a_4+2a_6)(x-1)^7$$
and $a_4,a_6\in \mathbb{F}_{3^m}$ arbitrary.

\par
  \item{\textsl{Case 2.}}
  $k=2\nu$ where $1\leq\nu\leq \frac{3^2-1}{4}=2$. If $\nu=2$, there is $1$ codes:

\par
  $\bullet$ $\langle u(x-1)^4,(x-1)^5\rangle$.

\par
  If $\nu=1$, we have $3^2-1-4\nu=4$, $\frac{3^2-1}{2}-2\nu=2$ and
$2=\frac{3^2-1}{4}-\nu+1\leq j\leq \frac{3^2-1}{2}-2\nu+1=2$. Hence
$$\Upsilon_{3}^{[2;4)}=\Upsilon_{2\cdot2-1}^{[2;4)}=\left(\begin{array}{c} 2\cr 0\end{array}\right)
\ {\rm and} \left(\begin{array}{c} b_2\cr b_3\end{array}\right)=a_2\Upsilon_{3}^{[2;4)}
=\left(\begin{array}{c} 2a_2\cr 0\end{array}\right).$$
In this case, there are $3^m$ codes:

\par
  $\bullet$ $\langle (x-1)^3b(x)+u(x-1)^2,(x-1)^7\rangle$, where $b(x)=2a_2(x-1)^2$ and $a_2\in \mathbb{F}_{3^m}$ arbitrary.

\par
 \item{\textsl{Case 3.}}
  $k=2\nu-1$ where $1\leq\nu\leq 2$. If $\nu=2$, there is $1$ codes:

\par
  $\bullet$ $\langle u(x-1)^3,(x-1)^6\rangle$.

\par
  If $\nu=1$, we have $3^2-1-4\nu+2=6$, $\frac{3^2-1}{2}-2\nu+1=3$ and
$3=\frac{3^2-1}{4}-\nu+2\leq j\leq \frac{3^2-1}{2}-2\nu+1=3$. Hence
$$\Upsilon_{5}^{[3;6)}=\Upsilon_{2\cdot3-1}^{[3;6)}=\left(\begin{array}{c} 0 \cr 2\cr 1\end{array}\right)
\ {\rm and} \left(\begin{array}{c} b_3\cr b_4 \cr b_5\end{array}\right)=a_4\Upsilon_{5}^{[3;6)}
=\left(\begin{array}{c} 0 \cr 2a_4\cr a_4\end{array}\right).$$
In this case, there are $3^m$ codes:

\par
  $\bullet$ $\langle (x-1)^2b(x)+u(x-1),(x-1)^8\rangle$, where $b(x)=2a_4(x-1)^4+a_4(x-1)^5$ and $a_4\in \mathbb{F}_{3^m}$
   arbitrary.
\end{description}

\par
  {\bf (III)} Let $s=3$. Then $3^3\equiv 3$ (mod 4) and $\frac{3^3+1}{4}=7$. In this case, the number
of self-dual cyclic codes over $\mathbb{F}_{3^m}+u\mathbb{F}_{3^m}$ of length $3^3$ is
$${\rm NE}(\mathbb{F}_{3^m}+u\mathbb{F}_{3^m},3^2)=2\cdot\frac{(3^m)^7-1}{3^m-1}
=2\sum_{l=0}^6(3^m)^l.$$
by Theorem 3($\dag$). From
$G_{27}+I_{27}=\left(\begin{array}{ccc} G_9+I_9 & 0 & 0\cr -G_9 & -G_9+I_9 & 0 \cr G_9 & -G_9 & G_9+I_9\end{array}\right)$, we list all these codes by the following two cases:

\begin{description}
\item{\textsl{Case 1.}}
   $k=2\nu$ where $0\leq \nu\leq 6$. In this case, there are $3^{6m}+3^{5m}+3^{4m}+3^{3m}+3^{2m}+3^m+1$
self-dual cyclic codes over $\mathbb{F}_{3^m}+u\mathbb{F}_{3^m}$ of length $3^3$ given by:

\par
  $\bullet$ $\langle u(x-1)^{12}, (x-1)^{15}\rangle$.

\par
  $\bullet$ $\langle (x-1)^{11}b(x)+u(x-1)^{10}, (x-1)^{17}\rangle$, \\
 where $b(x)=b_3(x-1)^3+b_4(x-1)^4+b_5(x-1)^5$ with
$$(b_3,b_4,b_5)=\left(a_4\Upsilon_{5}^{[3,6)}\right)^{{\rm tr}}=a_4(0,2,1)=(0,2a_4,a_4),$$
and $a_4\in \mathbb{F}_{3^m}$ arbitrary.

\par
  $\bullet$ $\langle (x-1)^{9}b(x)+u(x-1)^{8}, (x-1)^{19}\rangle$,
  where $b(x)=b_5(x-1)^5+b_6(x-1)^6+b_7(x-1)^7+b_8(x-1)^8+b_9(x-1)^9$ with
\begin{eqnarray*}
(b_5,b_6,b_7,b_8,b_9)&=&\left(a_6\Upsilon_{7}^{[5,10)}+a_8\Upsilon_{9}^{[5,10)}\right)^{{\rm tr}}\\
 &=&(a_6,a_8)\left(\Upsilon_{7}^{[5,10)},\Upsilon_{9}^{[5,10)}\right)^{{\rm tr}}\\
 &=&(a_6,a_8)\left(\begin{array}{ccccc}0 & 2 &-1 & 1 & 0 \cr & & 0 & 2 & 0\end{array}\right)\\
 &=&a_6(0,2,-1,1,0)+a_8(0,0,0,2,0)\\
 &=&(0,2a_6,2a_6,a_6+2a_8,0),
\end{eqnarray*}
and $a_6,a_8\in \mathbb{F}_{3^m}$ arbitrary.

\par
  $\bullet$ $\langle (x-1)^{7}b(x)+u(x-1)^{6}, (x-1)^{21}\rangle$,
  where $b(x)=\sum_{i=7}^{13}b_i(x-1)^i$ with
\begin{eqnarray*}
 &&(b_7,b_8,b_9,b_{10},b_{11},b_{12},b_{13})\\
 &=&(a_8,a_{10},a_{12})\left(\Upsilon_{9}^{[7,14)},\Upsilon_{11}^{[7,14)},\Upsilon_{13}^{[7,14)}\right)^{{\rm tr}}\\
 &=&(a_8,a_{10},a_{12})\left(\begin{array}{ccccccc}
 0 & 2 & 0 & 0 & 0 & 0 & 0 \cr
   &   & 0 & 2 & 1 & 0 & -1\cr
   &   &   &   & 0 & 2 & -1 \end{array}\right)\\
&=&(0,2a_8,0,2a_{10},a_{10},2a_{12},2a_{10}+2a_{12}),
\end{eqnarray*}
and $a_8,a_{10},a_{12}\in \mathbb{F}_{3^m}$ arbitrary.

\par
  $\bullet$ $\langle (x-1)^{5}b(x)+u(x-1)^{4}, (x-1)^{23}\rangle$,
  where $b(x)=\sum_{i=9}^{17}b_i(x-1)^i$ with
\begin{eqnarray*}
 &&(b_9,b_{10},b_{11},b_{12},b_{13},b_{14},b_{15},b_{16},b_{17})\\
 &=&(a_{10},a_{12},a_{14},a_{16})\left(\Upsilon_{11}^{[9,18)},\Upsilon_{13}^{[9,18)},\Upsilon_{15}^{[9,18)},
 \Upsilon_{17}^{[9,18)}\right)^{{\rm tr}}\\
 &=&(a_{10},a_{12},a_{14},a_{16})\left(\begin{array}{ccccccccc}
     0 & 2 & 1 & 0 & -1 & -1 & 0 & 1  & 1 \cr
       &   & 0 & 2 & -1 & 1  & 1 & -1 & 1 \cr
       &   &   &   & 0  & 2  & 0 & 0  & 1 \cr
        &   &   &   &   &    & 0 & 2  & 1
 \end{array}\right)\\
 &=&(0,2a_{10},a_{10},2a_{12},2a_{10}+2a_{12},2a_{10}+a_{12}+2a_{14},a_{12},\\
 && a_{10}+2a_{12}+2a_{16},a_{10}+a_{12}+a_{14}+a_{16}),
\end{eqnarray*}
and $a_{10},a_{12},a_{14},a_{16}\in \mathbb{F}_{3^m}$ arbitrary.

\par
  $\bullet$ $\langle (x-1)^{3}b(x)+u(x-1)^{2}, (x-1)^{25}\rangle$,
  where $b(x)=\sum_{i=11}^{21}b_i(x-1)^i$ with
{\small
\begin{eqnarray*}
 &&(b_{11},b_{12},b_{13},b_{14},b_{15},b_{16},b_{17},b_{18},b_{19},b_{20},b_{21})\\
 &=&(a_{12},a_{14},a_{16},a_{18},a_{20})\left(\Upsilon_{13}^{[11,22)},\Upsilon_{15}^{[11,22)},\Upsilon_{17}^{[11,22)},
   \Upsilon_{19}^{[11,22)},\Upsilon_{21}^{[11,22)}\right)^{{\rm tr}}\\
 &=& (a_{12},a_{14},a_{16},a_{18},a_{20})\left(\begin{array}{ccccccccccc}
     0 & 2 & -1 & 1 & 1 & -1 & 1 & 0 & 0 & 0 & 1 \cr
       &   & 0  & 2 & 0 & 0 &  1 & 0 & 0 & 0 & 0 \cr
       &   &    &   & 0 & 2 &  1 & 0 & 0 & 0 & 0 \cr
       &   &    &   &   &   &  0 & 2 &-1 & 1 & -1\cr
       &   &    &   &   &   &    &   & 0 & 2 & 0
 \end{array}\right)\\
 &=&(0,2a_{12},2a_{12},a_{12}+2a_{14},a_{12},2a_{12}+2a_{16},a_{12}+a_{14}+a_{16},2a_{18},\\
 && 2a_{18},a_{18}+2a_{20},a_{12}+2a_{18}),
\end{eqnarray*}  }
and $a_{12},a_{14},a_{16},a_{18},a_{20}\in \mathbb{F}_{3^m}$ arbitrary.

\par
  $\bullet$ $\langle (x-1)b(x)+u\rangle$,
  where $b(x)=\sum_{i=13}^{25}b_i(x-1)^i$ with
\begin{eqnarray*}
 &&(b_{13},b_{14},b_{15},b_{16},b_{17},b_{18},b_{19},b_{20},b_{21},b_{22},b_{23},b_{24},b_{25})\\
 &=&\left(\sum_{j=8}^{13}a_{2j-2}\Upsilon_{2j-1}^{[13,26)}\right)^{{\rm tr}}\\
 &=&A\left(\begin{array}{ccccccccccccc}
    0 & 2 & 0 & 0 & 1 & 0 & 0 & 0 & 0 & 0 & 1 & 0 & 0 \cr
      &   & 0 & 2 & 1 & 0 &  0 & 0 & 0  & 0 & 0  & 0 & 1 \cr
      &   &   &   & 0 & 2 & -1 & 1 & -1 & 1 & -1 & 1 & -1 \cr
      &   &   &   &   &   & 0  & 2 & 0  & 0 & -1 & 0 & 0 \cr
      &   &   &   &   &   &    &   & 0  & 2 & 1 & 0 & 1  \cr
      &   &   &   &   &   &    &   &    &   & 0 & 2 & -1
 \end{array}\right)\\
 &=&(0,2a_{14},0,2a_{16},a_{14}+a_{16},2a_{18},2a_{18},a_{18}+2a_{20},2a_{18},a_{18}+2a_{22},\\
  &&
   a_{14}+2a_{18}+2a_{20}+a_{22},a_{18}+2a_{24}, a_{16}+2a_{18}+a_{22}+2a_{24}),
\end{eqnarray*}
and $A=(a_{14},a_{16},a_{18},a_{20},a_{22},a_{24})\in \mathbb{F}_{3^m}^6$ arbitrary.

\item{\textsl{Case 2.}}
   $k=2\nu+1$ where $0\leq \nu\leq 6$. In this case, there are $3^{6m}+3^{5m}+3^{4m}+3^{3m}+3^{2m}+3^m+1$
self-dual cyclic codes over $\mathbb{F}_{3^m}+u\mathbb{F}_{3^m}$ of length $3^3$ given by:

\par
  $\bullet$ $\langle u(x-1)^{13}, (x-1)^{14}\rangle$.

\par
  $\bullet$ $\langle (x-1)^{12}b(x)+u(x-1)^{11}, (x-1)^{16}\rangle$, where
$b(x)=b_2(x-1)^2+b_3(x-1)^3$ with
$$(b_2,b_3)=a_2(\Upsilon_{3}^{[2,4)})^{{\rm tr}}=(2a_2,0), \
{\rm and} \ a_2\in \mathbb{F}_{3^m} \ {\rm arbitrary}.$$

\par
  $\bullet$ $\langle (x-1)^{10}b(x)+u(x-1)^{9}, (x-1)^{18}\rangle$, where
$b(x)=b_4(x-1)^4+b_5(x-1)^5+b_6(x-1)^6+b_7(x-1)^7$ with
\begin{eqnarray*}
 (b_4,b_5,b_6,b_7)
 &=&(a_4,a_6)\left(\Upsilon_{5}^{[4,8)},\Upsilon_{7}^{[4,8)}\right)^{{\rm tr}}
 =(a_4,a_6)\left(\begin{array}{cccc} 2 & 1 & 0 & 1\cr & & 2 & -1\end{array}\right)\\
 &=&(2a_4,a_4,2a_6,a_4+2a_6),
\end{eqnarray*}
and $a_4,a_6\in \mathbb{F}_{3^m}$ arbitrary.

\par
  $\bullet$ $\langle (x-1)^{8}b(x)+u(x-1)^{7}, (x-1)^{20}\rangle$, where
$b(x)=b_6(x-1)^6+b_7(x-1)^7+b_8(x-1)^8+b_9(x-1)^9+b_{10}(x-1)^{10}+b_{11}(x-1)^{11}$ with
\begin{eqnarray*}
(b_6,b_7,b_8,b_9,b_{10},b_{11})
 &=&(a_6,a_8,a_{10})\left(\Upsilon_{7}^{[6,12)},\Upsilon_{9}^{[6,12)},\Upsilon_{11}^{[6,12)}\right)^{{\rm tr}}\\
 &=&(a_6,a_8,a_{10})\left(\begin{array}{cccccc} 2 & -1 & 1 & 0 & 0 & 0\cr & & 2 & 0 & 0 & 0 \cr & & & & 2 & 1\end{array}\right)\\
 &=&(2a_6,2a_6,a_6+2a_8,0,2a_{10},a_{10}),
\end{eqnarray*}
and $a_6,a_8,a_{10}\in \mathbb{F}_{3^m}$ arbitrary.

\par
  $\bullet$ $\langle (x-1)^{6}b(x)+u(x-1)^{5}, (x-1)^{22}\rangle$, where
$b(x)=\sum_{i=8}^{15}(x-1)^i$ with
\begin{eqnarray*}
 &&(b_8,b_9,b_{10},b_{11},b_{12},b_{13},b_{14},b_{15})\\
 &=&(a_8,a_{10},a_{12},a_{14})\left(\Upsilon_{9}^{[8,16)},\Upsilon_{11}^{[8,16)},\Upsilon_{13}^{[8,16)},
  \Upsilon_{15}^{[8,16)}\right)^{{\rm tr}}\\
 &=&(a_8,a_{10},a_{12},a_{14})\left(\begin{array}{cccccccc}
 2 & 0 & 0 & 0 & 0 & 0 & 0 & 0\cr & & 2 & 1 & 0 & -1 & -1 & 0 \cr & & & & 2 & -1 & 1 & 1\cr
 & & & & & & 2 & 0\end{array}\right)\\
 &=&(a_8,0,2a_{10},a_{10},2a_{12},2a_{10}+2a_{12},2a_{10}+a_{12}+2a_{14},a_{12}),
\end{eqnarray*}
and $a_8,a_{10},a_{12},a_{14}\in \mathbb{F}_{3^m}$ arbitrary.

\par
  $\bullet$ $\langle (x-1)^{4}b(x)+u(x-1)^{3}, (x-1)^{24}\rangle$, where
$b(x)=\sum_{i=10}^{19}(x-1)^i$ with
\begin{eqnarray*}
 &&(b_{10},b_{11},b_{12},b_{13},b_{14},b_{15},b_{16},b_{17},b_{18},b_{19})\\
 &=&(a_{10},a_{12},a_{14},a_{16},a_{18})\left(\Upsilon_{11}^{[10,20)},\Upsilon_{13}^{[10,20)},
  \Upsilon_{15}^{[10,20)},\Upsilon_{17}^{[10,20)},\Upsilon_{19}^{[10,20)}\right)^{{\rm tr}}\\
 &=&(a_{10},a_{12},a_{14},a_{16},a_{18})\left(\begin{array}{cccccccccc}
 2 & 1 & 0 & -1 & -1 & 0 & 1 & 1 & 0 & 1\cr & & 2 & -1 & 1 & 1 & -1 & 1 & 0 & 0 \cr & & & & 2 & 0 & 0 & 1 & 0 & 0\cr
 & & & & & & 2 & 1 & 0 & 0\cr  & & & & & &  &  & 2 &-1\end{array}\right)\\
 &=&(2a_{10},a_{10},2a_{12},2a_{10}+2a_{12},2a_{10}+a_{12}+2a_{14},a_{12},\\
 && a_{10}+2a_{12}+2a_{16},a_{10}+a_{12}+a_{14}+a_{16},2a_{18},a_{10}+2a_{18}),
\end{eqnarray*}
and $a_{10},a_{12},a_{14},a_{16},a_{18}\in \mathbb{F}_{3^m}$ arbitrary.

\par
  $\bullet$ $\langle (x-1)^{2}b(x)+u(x-1), (x-1)^{26}\rangle$, where
$b(x)=\sum_{i=12}^{23}(x-1)^i$ with
\begin{eqnarray*}
 &&(b_{12},b_{13},b_{14},b_{15},b_{16},b_{17},b_{18},b_{19},b_{20},b_{21},b_{22},b_{23})\\
 &=&A\left(\Upsilon_{13}^{[12,24)},
  \Upsilon_{15}^{[12,24)},\Upsilon_{17}^{[12,24)},\Upsilon_{19}^{[12,24)},\Upsilon_{21}^{[12,24)},
  \Upsilon_{23}^{[12,24)}\right)^{{\rm tr}}\\
 &=&A\left(\begin{array}{cccccccccccc}
 2 & -1 & 1 & 1 & -1 & 1 & 0 & 0 & 0 & 1 & -1 & 1\cr & & 2 & 0 & 0 & 1 & 0 & 0 & 0 & 0 & 0 & 1
 \cr & & & & 2 & 1 & 0 & 0 & 0 & 0 & 0 & 0\cr
 & & & & & & 2 & -1 & 1 & -1 & 1 & -1\cr  & & & & & &  &  & 2 & 0 & 0 & -1
 \cr   & & & & & &  &  & & & 2 & 1\end{array}\right)\\
 &=& (2a_{12},2a_{14},a_{12}+2a_{14},a_{12},2a_{12}+2a_{16},a_{12}+a_{14}+a_{16},2a_{18},2a_{18},\\
 && a_{18}+2a_{20},a_{12}+2a_{18},2a_{12}+a_{18}+2a_{22},\\
 && a_{12}+a_{14}+2a_{18}+2a_{20}+a_{22}),
\end{eqnarray*}
and $A=(a_{12},a_{14},a_{16},a_{18},a_{20},a_{22})\in \mathbb{F}_{3^m}^6$ arbitrary.

\end{description}

\section{Conclusions and further research}
\noindent
For any odd prime number $p$ and positive integers $m,s$, we have given an explicit representation for all distinct self-dual cyclic codes of length $p^s$
over the finite chain ring $\mathbb{F}_{p^m}
+u\mathbb{F}_{p^m}$ ($u^2=0$) by a new way different from that used in \cite{s5} and \cite{s7}. In particular,  we provide an efficient method to construct precisely  all distinct self-dual cyclic codes of length $p^s$
over $\mathbb{F}_{p^m}+u\mathbb{F}_{p^m}$ by use of Kronecker products
of matrices over $\mathbb{F}_p$ with a specific type.

\par
   Giving an explicit
representation and enumeration for self-dual cyclic codes and self-dual negacyclic codes
over $\mathbb{F}_{p^m}+u\mathbb{F}_{p^m}$ for arbitrary length and obtaining some bounds for the minimal distance such as BCH-like of a self-dual cyclic code over the ring $\mathbb{F}_{p^m}+u\mathbb{F}_{p^m}$ by just looking at its representation of such codes are future topics of interest.

\section*{Acknowledgments}

\noindent
Part of this work was done when Yonglin Cao was visiting Chern Institute of Mathematics,
Nankai University, Tianjin, China. Yonglin Cao would like to thank the institution for the kind hospitality. This research is
supported in part by the National Natural Science Foundation of
China (Grant Nos. 11671235, 11801324),  the Shandong Provincial Natural Science Foundation, China
(Grant No. ZR2018BA007), the Scientific Research Fund of Hubei Provincial Key Laboratory of Applied Mathematics (Hubei University)
(Grant No. AM201804) and the Scientific Research Fund of Hunan
Provincial Key Laboratory of Mathematical Modeling and Analysis in
Engineering (No. 2018MMAEZD09). H.Q. Dinh is
grateful for the Centre of Excellence in Econometrics, Chiang Mai University,
Thailand, for partial financial support.

% You may incorporate your references as follows in your main tex file.
% Using BibTex is not recommended but can be handled.

\end{document}